\documentclass[twocolumn]{aastex631}
\usepackage{mathrsfs}
\usepackage{tikz}
\usetikzlibrary{positioning}
\usepackage{subfigure}
\usepackage{graphicx}
\usepackage{epstopdf}
\usepackage{epsfig}
\usepackage{overpic}
\usepackage{ulem}
\usepackage{soul}
\soulregister\citep7
\soulregister\ref7
\soulregister\citealt7
\soulregister\cite7
\usepackage{float}
\usepackage{lineno}

\makeatletter

\newcommand{\Rmnum}[1]{\expandafter\@slowromancap\romannumeral #1@}
\makeatother

\shorttitle{Simulation of AR 10930 eruption}
\shortauthors{Wang et al.}

\graphicspath{{./}{figures/}}

\begin{document}

\title{MHD Simulation of Homologous Eruptions from Solar Active Region 10930 Caused by Sunspot Rotation}

\author{Xinyi Wang}
\affiliation{SIGMA Weather Group, State Key Laboratory for Space Weather, National Space Science Center, Chinese Academy of Sciences,Beijing 100190, PR China}
\affiliation{College of Earth and Planetary Sciences, University of Chinese Academy of Sciences, Beijing 100049, PR China}

\author[0000-0002-7018-6862]{Chaowei Jiang}
\affiliation{Institute of Space Science and Applied Technology, Harbin Institute of Technology, Shenzhen 518055, PR China, \url{chaowei@hit.edu.cn}}

\author[0000-0001-8605-2159]{Xueshang Feng}
\affiliation{College of Earth and Planetary Sciences, University of Chinese Academy of Sciences, Beijing 100049, PR China}
\affiliation{Institute of Space Science and Applied Technology, Harbin Institute of Technology, Shenzhen 518055, PR China, \url{chaowei@hit.edu.cn}}

\author[0000-0002-1916-1053]{Aiying Duan}
\affiliation{Planetary Environmental and Astrobiological Research Laboratory (PEARL), School of Atmospheric Sciences, Sun Yat-sen University, Zhuhai 519000, PR China}

\author{Xinkai Bian}
\affiliation{Institute of Space Science and Applied Technology, Harbin Institute of Technology, Shenzhen 518055, PR China, \url{chaowei@hit.edu.cn}}

\begin{abstract}

The relationship between solar eruption and sunspot rotation has been widely reported, and the underlying mechanism requires to be studied. Here we performed a full 3D MHD simulation of data-constrained approach to study the mechanism of flare eruptions in active region (AR) NOAA 10930, which is characterized by continuous sunspot rotation and homologous eruptions. We reconstructed the potential magnetic field from the magnetogram of Hinode/SOT as the initial condition and drove the MHD system by applying continuous sunspot rotation at the bottom boundary. The key magnetic structure before the major eruptions and the pre-formed current sheet were derived, which is responsible for the complex MHD evolution with multiple stages. The major eruptions were triggered directly by fast reconnection in the pre-formed current sheet above the main polarity inversion line between the two major magnetic polarities of the AR. Furthermore, our simulation shows the homologous eruption successfully. It has reasonable consistence with observations in relative strength, energy release, X-ray and $\rm H\alpha$ features and time interval of eruptions. In addition, the rotation angle of the sunspot before the first eruption in the simulation is also close to the observed value. Our simulation offers a scenario different from many previous studies based on ideal instabilities of twisted magnetic flux rope, and shows the importance of sunspot rotation and magnetic reconnection in efficiently producing homologous eruptions by continuous energy injection and impulsive energy release in a recurrent way.

\end{abstract}


\section{Introduction}
Solar eruption is considered as the most magnificent phenomenon in the solar system. It is manifested as flares and coronal mass ejections along with high energetic particle events, and these solar transients from solar corona can affect heavily the solar-terrestrial environment. By estimating the typical parameters of the eruptive source regions, it has been well recognized that only the magnetic free energy stored in the coronal current exceeds the required energy density as released in a typical eruption \citep{2000JGR10523153F}. Based on this, many theories of solar eruption have been proposed, which converged into the standard CSHKP model \citep{1964NASSP..50..451C, 1966Natur.211..695S, 1974SoPh...34..323H, 1976SoPh...50...85K} by grasping the key structure of magnetic field and can be applied to incorporate many observations (such as flare, particle acceleration, shock wave and radio burst).

 Although the basic scenario is well established, the initiation mechanism of solar eruption remains not fully understood. Currently two kinds of initiation mechanisms are frequently invoked, one is based on the ideal plasma macro-instabilities and the other on non-ideal, micro process, i.e., magnetic reconnection \citep{2011LRSP....8....1C}. The coronal magnetic field  in the non-eruptive evolution is nearly force-free, and a particular force-free structure, magnetic flux rope (MFR), holds the central position in models based on ideal instabilities. In the earliest model of such kind, the MFR is simply taken as an electric wire as in \cite{2000JGR...105.2375L} (later an MFR has a twisted 3D structure), which is confined to be in equilibrium by the overlying magnetic arcade that is anchored at the photosphere. The ideal instabilities of such a pre-existing MFR, mainly the torus and kink instability, provide an effective way for driving eruptions as developed by many theoretical and simulation researches \citep{1978mit..book.....B,2004A&A...413L..27T,2005ApJ...630..543F,2006PhRvL..96y5002K}. When the MFR reaches the critical point of instability, the eruption is triggered with a quick lifting up of the MFR. Meanwhile a current sheet (CS) forms under the erupting MFR in a dynamic way. A flare is resulted when magnetic reconnection sets in at the CS, which converts magnetic energy to thermal and nonthermal energies that power the flare. Usually the torus instability is considered to be more efficient, while the kink one can only bring MFRs to be torus unstable or, otherwise, confined flares \citep{2005ApJ...630L..97T,2013AdSpR..51.1967S}.

 The second type of mechanisms as built upon magnetic reconnection needs a CS forms before eruption, such as the bipolar tether-cutting model \citep{1980IAUS...91..207M,1992LNP...399...69M,2001ApJ...552..833M} and the quadrupolar breakout model \citep{1999ApJ...510..485A,2012ApJ...760...81K}. In these models, the CS first forms, and reconnection then triggers the eruption, while MFR forms during the eruption, which is distinct from the first type in which the CS is built up at the wake of the erupting MFR. In the breakout model, a magnetic null point needs to be pre-existing above a sheared core. The expansion of the sheared core will compress the null point to form the breakout CS. Slow reconnection in the CS progressively weakens the overlying field, which in turn allows more expansion of the core field, and it is proposed that a positive feedback is established, which finally leads to the formation of the flare CS (i.e., the vertical CS within the sheared core) and an eruption \citep{2012ApJ...760...81K}. The tether-cutting model is simpler in the requirement of magnetic topology, since it is based on only a bipolar arcade. Shearing motion in bipolar field near the polarity inversion line (PIL) forms a CS. The reconnection in the CS cuts gradually the tethering field lines, which allows the expansion of the core field until a global disruption of the system is triggered, similar to that of the breakout model. The reconnection in a newly-formed large-scale CS, as resulted from stretching of the large-scale overlying field by the rising core field (as an MFR), further plays an important role in supporting the eruption. However, previous numerical simulations (e.g., \citealt{2003ApJ...585.1073A,2010ApJ...708..314A}) show that shearing motion alone can only help to form an MFR (along with flux cancellation), while its eruption is triggered by some ideal instabilities. Recently a ultra-high accuracy MHD simulation has established a new point: the sling-shot effect of reconnection can accelerate impulsively the plasma to fast eruption without ideal instabilities taking place \citep{2021NatAs...5.1126J}, thus emphasizing the key role of reconnection in both triggering and driving a eruption. In the above models, magnetic energy ought to be released mainly by fast reconnection \citep{1964NASSP..50..425P} even in the presence of an erupting MFR. Ideal instabilities alone can only release a small amount of magnetic energy \citep{1991ApJ...373..294F}, which is inadequate for accelerating the coronal plasma.

 Even though so many theoretical models exist, it is not easy to determine which one operates in the realistic events, since the coronal magnetic fields and their evolutions associated with eruptions are often much more complex than as described in the models. In the recent years, numerical models that are constrained or directly driven by the observed data have been developed and proved to be a powerful tool in probing the mechanisms of realistic solar eruptions
(e.g., \citealt{2014ApJ...788..182I,2017ApJ...840...37P,2018ApJ...866...96J,2018shin.confE.208J,2021ApJ...919...39G}), and the progresses have been reviewed by \cite{JIANG2022100236}.

In this paper we performed an MHD simulation of a data-constrained approach to study the mechanism of the flare eruptions in active region (AR) NOAA 10930. This AR is very eruption-productive and has been studied extensively in many previous papers (e.g., \citealt{2007PASJ...59S.785S,2008ApJ...676L..81J,2010AGUFMSH31A1785I,2011ApJ...740...19R,2011ApJ...740...68F,2014Natur.514..465A}). It appeared on the solar disk on December 2006 and produced a number of flares including four X-class ones in a few days (e.g., an X3.4 flare on December 13th and an X1.9 flare on December 14th). The most prominent dynamics of the AR is that a sunspot newly emerging into the AR showed continual rotation for days in the period with the flares. For example, previous studies found that the sunspot had rotated about $240^{\circ}$ in 2 days \citep{2007ApJ...662L..35Z} or $540^{\circ}$ in 5 days \citep{2009SoPh..258..203M} as measured by different methods. Such rotation resulted in a strong shearing flow near the main PIL of the AR and the magnitude of the shearing speed has also been estimated \citep{2008PASJ...60.1181M,2009ApJ...690.1820T}. There are some static modeling of the coronal magnetic field for this AR  \citep{2008ApJ...675.1637S,2008ApJ...679.1629G}, i.e., by using nonlinear force-free field (NLFFF) extrapolation, verifying that the magnetic structure has a highly sheared core or MFR. A few works have been done also using dynamic MHD simulations with focus on the mechanism of eruptions, but the conclusions are at odds with each other. For instance, based on the observed vector magnetograms from Hinode/SOT \citep{2008SoPh..249..167T}, \cite{2014Natur.514..465A} first reconstructed a series of NLFFF solutions to follow the pre-flare evolution of the AR from December 9th to December 12th, and found that a sigmoidal MFR was progressively built up. Then with the pre-flare NLFFF solutions as the initial condition, they managed to simulate the eruption of the X3.4 flare by using 3 different types of ad-hoc boundary conditions at the bottom surface and concluded that the main trigger of the flare is torus instability of the MFR. \cite{2011ApJ...740...68F} constructed a background potential field by the line-of-sight magnetogram from SOHO/MDI for this AR and then introduced into the core field an artificial MFR through rigid emergence from the bottom boundary to simulate how the emerging MFR leads to the eruption in the background field. In such a study, they suggested a different trigger, i.e., kink instability rather than the torus instability because the decay index near the erupting MFR is found to be smaller than the critical value ($n<1.5$). Another study \citep{2017ApJ...842...86M} was performed by triggering the eruption with a small bipole emerging at the main PIL in a large-scale stable NLFFF field. By adjusting the orientations of the small bipole relative to the main PIL, they concluded that the so-called opposite polarity and reversed shear types' emergence can effectively trigger the eruption, as the same mechanism originally developed in \cite{2012ApJ...760...31K}.

We note that none of the aforementioned dynamic simulations have taken into consideration the effect the continual and significant rotation of the AR's sunspot in leading to the eruptions, which, however, is strongly suggested by observations \citep{1909MNRAS..69..454E,2003SoPh..216...79B,2008MNRAS.391.1887Y,2018ApJ...856...79Y,2012ApJ...761...60V}. Although the preflare sheared magnetic structure is undoubtedly resulted by the sunspot rotation, there is no self-consistent model of such a process, and this is also the motivation of this paper. Here we employed an MHD model as driven by the sunspot rotation to follow the coronal magnetic evolution of AR 10930 from its energy slow accumulation to fast releasing process. We started the simulation with a potential magnetic field reconstructed from the observed magnetogram and then applied a rotational motion to the positive sunspot of the AR to mimic the observed rotation. With a continual rotational driving, our model displayed a full evolution from the initial potential field to two homologous eruptions (which may correspond to 2 X-class flares). We found that reconnection in a quasi-statically pre-formed CS triggered the homologous eruptions, which is consistent with a fundamental mechanism of solar eruption initiation as recently established \citep{2021NatAs...5.1126J, 2021arXiv211104984B, 2022ApJ...925L...7B}. Furthermore, our results of the coronal magnetic configuration have reasonable consistency with the observed soft X-ray and $\rm H\alpha$ features. Also the time interval and relative strength of the simulated eruptions are on the same scale of the quantities of eruptions as derived from observations. The mechanism in our research is different from other works that requires the pre-formed MFR and is initiated by ideal MHD instabilities. We also suggest that many homologous eruptions of rotational sunspots may be triggered by the same mechanism in this paper.

The paper is organized as follows. We first show the observation and data in Section $\ref{Data and Observation}$, then describe the model and method in Section $\ref{Model and Method}$. Simulation results are displayed in Section $\ref{Results}$ and finally we give discussion and conclusion in Section $\ref{Conclusion}$.
\section{Data and Observation}\label{Data and Observation}
The AR NOAA 10930 is highly dynamic in which 4 X$-$class flares were produced and 3 of them occurred on December, 2006; X6.5 on December 6th, X3.4 on December 13th and X1.5 on December 14th \citep{2007PASJ...59S.779K}. In this research we focus on the X3.4 flare located at S07W22 on December 13th and the X1.5 flare located at S06W46 on December 14th \citep{2013ApJ...778...48B}.

\begin{figure}[htbp]
	\centering  
	\includegraphics[scale=0.5]{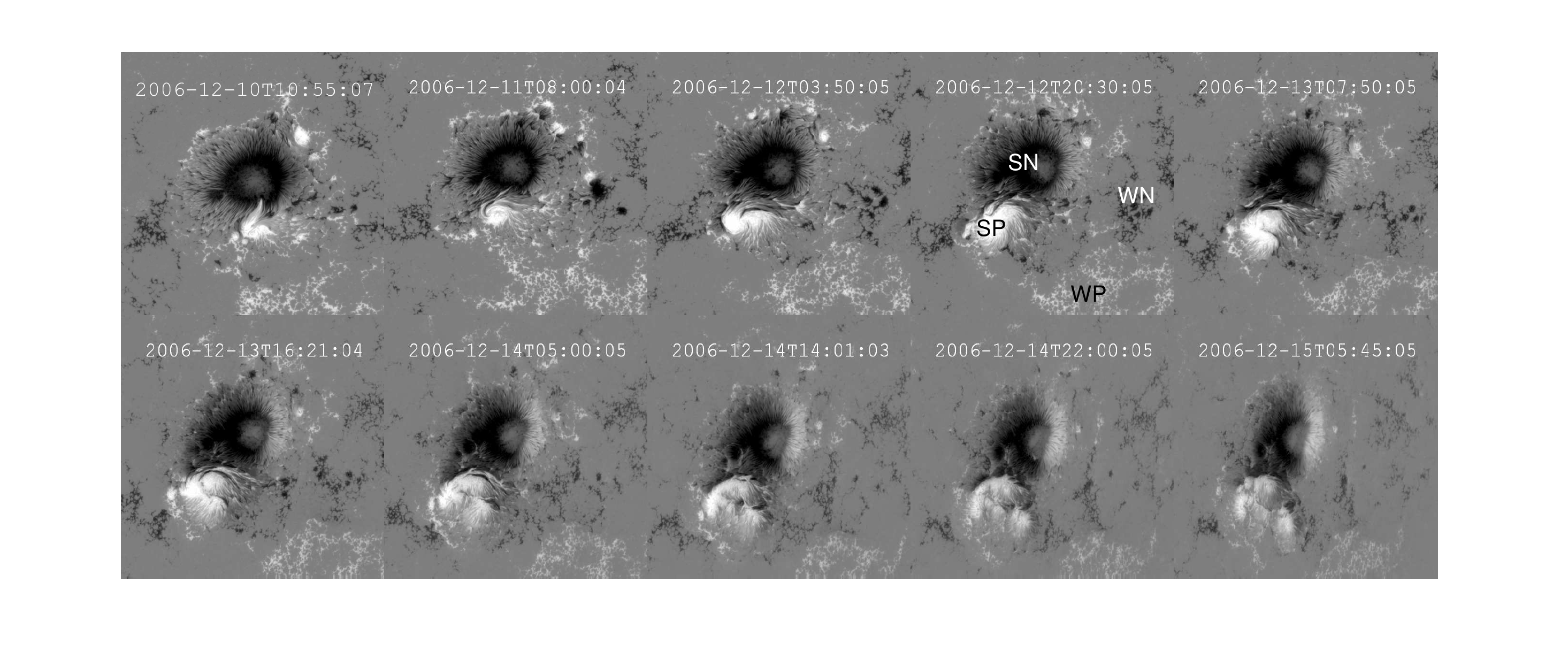}
	\caption{Time series of Stokes V images of Hinode/SOT from December 10th to December 15th. Black and white represents the negative and positive polarity respectively.}
	\label{sunspot}
\end{figure}

Figure~\ref{sunspot} taken from Stoke V images of HINODE/SOT \citep{2007SoPh..243....3K,2008SoPh..249..167T} shows the complex magnetic flux distribution of this AR. As denoted in the 4th panel of the figure, we define 4 areas of the photospheric magnetic field, which are, respectively, the strong positive (SP, which is the rotating sunspot), the weak positive (WP, the weak field region with positive polarities at the west side of the rotating sunspot), the strong negative (SN, i.e., the large sunspot) and the weak negative (WN, the weak field region of negative polarity at the west side of the large sunspot). There are mainly two sunspots with opposite magnetic polarities. The leading sunspot (SN) shows nearly no change during two flare events, while the smaller one in the south (SP) shows evident growth, and this indicates that the main sunspots are not a pair. The positive sunspot emerged later than the main negative sunspot, translating from west to east (right to left) \citep{2008ApJ...687..658W} with an obvious counter-clockwise rotation from December 10th to December 14th. It is connected not only to the main negative sunspot in the north but also to the west (right) dispersed polarities (WN) \citep{2009SoPh..258..203M}. The positive sunspot became diffused and rotated more slowly on December 14th.

\begin{figure}[htbp]
	\centering  
	\includegraphics[scale=0.5]{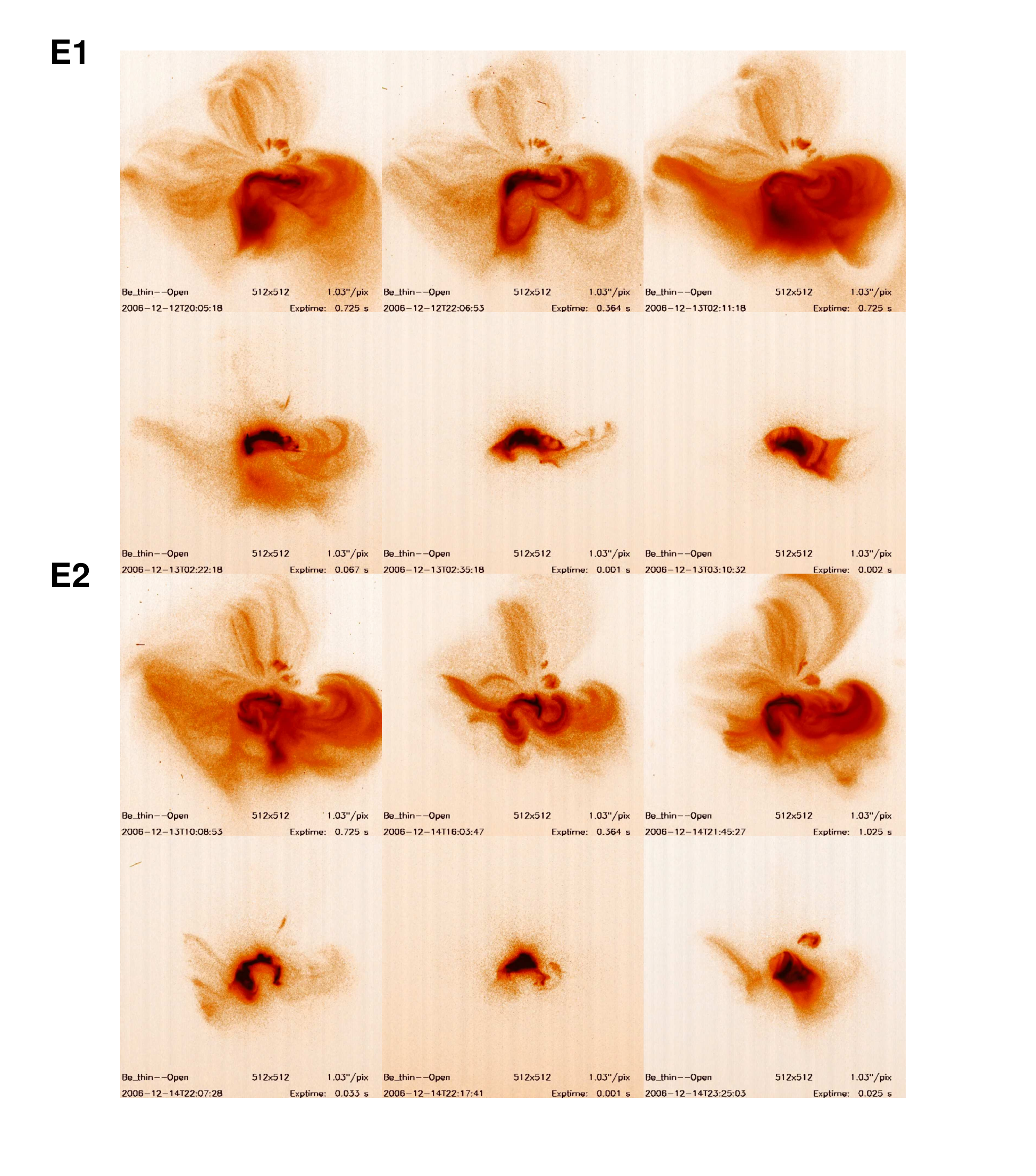}
	\caption{Time series of X-ray images of two X-class flares taken from Hinode/XRT.}
	\label{XRAY}
\end{figure}

\begin{figure}[htbp]
	\centering  
	\includegraphics[scale=0.5]{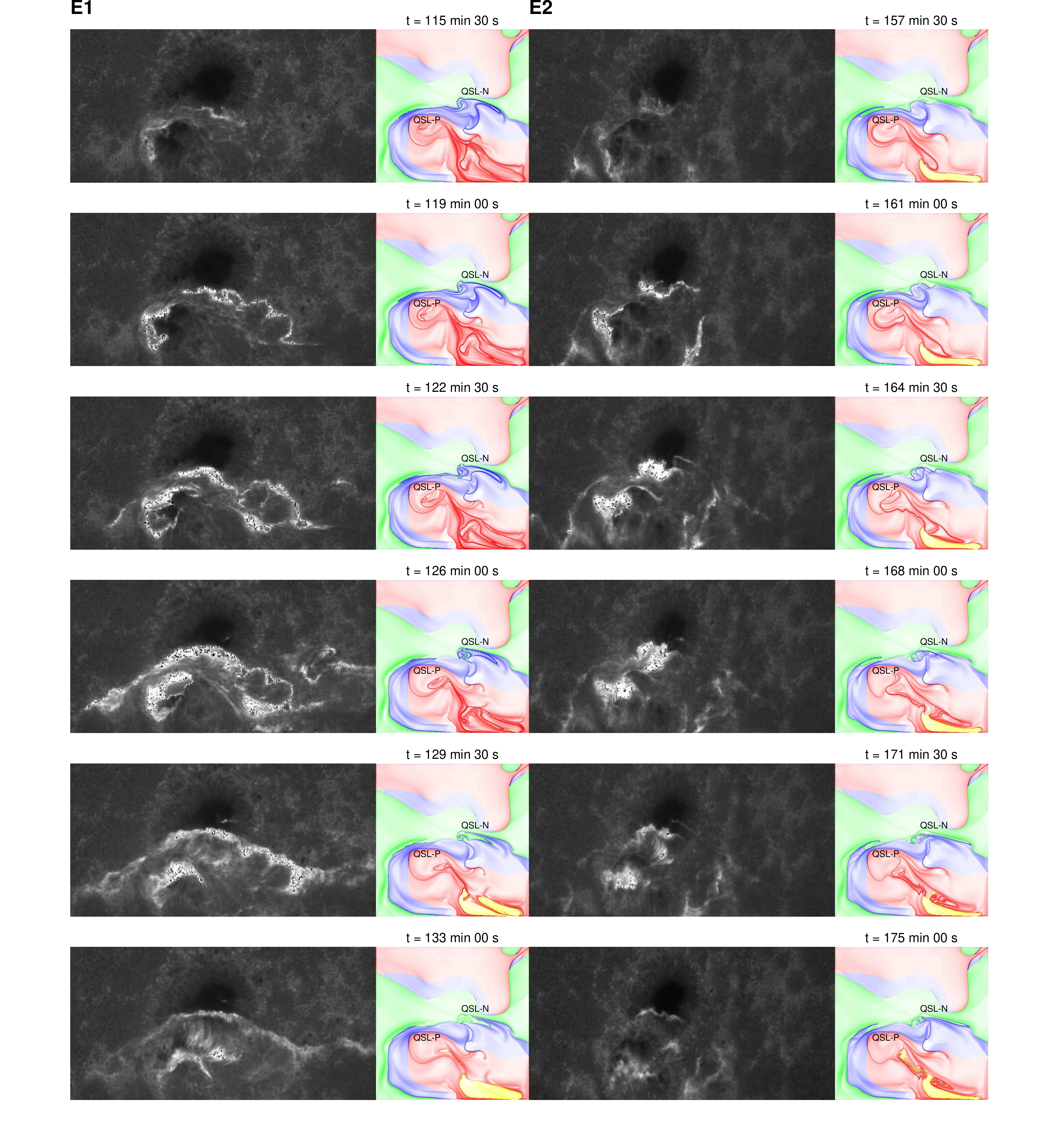}
	\caption{$\mathbf{E}$1): Two columns shows the $\rm H\alpha$ ribbons images taken from SOT/BFI and the bottom QSL evolution during E1, respectively. $\mathbf{E}$2): The same as E1 but during the second eruption. The green and yellow region of QSL are the open field of negative and positive field respectively. The red and blue region denotes the close field of positive and negative field. The upper limit of Q factor in red and blue region is $\rm log_{10}$$Q$$_{\rm max}=5$ }
	\label{QSLflare}
\end{figure}

 The evolution of an inverse-S sigmoid during two flares is shown in Figure~\ref{XRAY} taken from Hinode/XRT \citep{2005AdSpR..36.1489D,2007SoPh..243...63G}. The sigmoid formed near the main PIL and had a big tail around December 12th, which is likely formed by the rotation of the positive sunspot. The post-flare arcades spread from left to right during the first flare. In the middle of December 13th, the first flare ended, and the sigmoid reformed at the same position was involved in the second flare. The post-flare arcades during the second eruption didn't spread too much to the right as the flare ribbons. Both flares exhibited two ribbons signature as shown in the 1st and 3rd columns of Figure~\ref{QSLflare}, which are taken from Broadband Filter Imager (BFI) of SOT. The negative ribbon of both flares was located between the two main spots, spreading to right and was longer in the first eruption. While the positive ribbon was initially on the left side of the positive polarity (since the footpoints at the positive polarity have been rotated counter-clockwise to left) and it shrank into a circle to the right (Figure~\ref{QSLflare}). Except the length of the negative ribbon, the corresponding ribbons in both events resemble each other in both position and shape, which reflects the similarity in their underlying magnetic configurations and thus their trigger mechanisms. There was an Earth$-$directed CME with a projected speed of $\rm 1780~km~s^{-1}$ \citep{2010BASI...38..147R} on December 13th and another CME with a speed of $\rm 1042~km~s^{-1}$ on December 14th. A major geomagnetic storm was observed on December 15th.
\section{Model and Method}\label{Model and Method}
We used the DARE--MHD model \citep{2016NatCo...711522J} to study the dynamic evolution of the solar corona. The model was developed based on the CESE method \citep{2002JCoPh.175..168Z,1993LNP...414..396C} in Cartesian coordinate system combined with adaptive mesh refinement (AMR) technique by utilizing the PARAMESH \citep{2000CoPhC.126..330M} to solve the full MHD equations:
\begin{equation}\label{MHD equation}
\begin{array}{r}
\frac{\partial \rho}{\partial t}+\nabla \cdot(\rho \mathbf{v})=-\nu_{\rho}\left(\rho-\rho_{0}\right) \\
\rho \frac{D \mathbf{v}}{D t}=-\nabla p+\mathbf{J} \times \mathbf{B}+\rho \mathbf{g}+\nabla \cdot(\nu \rho \nabla \mathbf{v}) \\
\frac{\partial \mathbf{B}}{\partial t}=\nabla \times(\mathbf{v} \times \mathbf{B}-\eta \mu_{0} \mathbf{J}) \\
\frac{\partial T}{\partial t}+\nabla \cdot(T \mathbf{v})=(2-\gamma) T \nabla \cdot \mathbf{v}
\end{array}
\end{equation}
 where $\mathbf{J}=\nabla \times \mathbf{B}/\mu_{0}$, $\mathbf{g}$ is the solar gravity, $\mu_{0}$ is the magnetic permeability in vacuum, $\nu$ is the kinetic viscosity and $\gamma=1$ is the adiabatic index. We choose $\nu_{\rho}= 0.05~V_{A}$ (the $\rm Alfv \acute{e} n$ speed) to avoid the very low density in the strong magnetic field region, which may lead to a very small time step. By setting this, the plasma density will be relaxed to its initial value $\rho_{0}$ in a time scale of 20 $\rm Alfv \acute{e} n$ time $\tau_{\rm A}$. This time scale is sufficient large such that the fast dynamics of $\rm Alfv \acute{e} nic$ speed is not influenced. The viscosity $\nu$ is given as $\nu=0.05\Delta x^{2}/\Delta t$, where $\Delta x$ (varies from $1^{\prime \prime}$ to $4^{\prime \prime}$ in our simulation) and $\Delta t$ are the grid resolution and time step respectively. No explicit resistivity was applied in our simulation. Since to mimic the real corona environment, any explicit value of $\eta$ will give a larger value of resistivity than only a numerical method has, which will affect the reconnection process. The computational domain is sufficiently large of $\rm [-553,553]~Mm$ in $x$,$y$-direction and $\rm [0,1106]~Mm$ in $z$-direction to prevent the influence of the side and top boundary conditions on the computation of the eruption initiation. The Powell source terms and the diffusion control term are added to maintain the divergence-free condition of magnetic field as described in \cite{2010SoPh..267..463J}.
\subsection{Initial Conditions}
We smoothed the magnetogram at December 12th 2006, 20:30 UT which is taken from \cite{2008ApJ...675.1637S} using Gaussian smoothing with FWHM of 20 pixels. This makes the maximum value of $B_{z}$ decreases from 2619 G to 1595 G and then we constructed a potential field as the initial condition. The background plasma density satisfies a hydrostatic isothermal model with a value of $\rm 2.3\times 10^{-15}~g~cm^{-3}$ at bottom. To save the computation time, the strength of magnetic field from the smoothed magnetogram is reduced by a factor of 25. But this will make the plasma pressure and density decay slower than the background magnetic field, causing a higher plasma $\beta$, if we use the real value of solar gravity ($g_{\odot}=\rm 274~m~s^{-2}$). To avoid such a situation, we modified the gravity in the same way as \cite{2021NatAs...5.1126J}:
\begin{equation}\label{modified gravity}
g=\frac{k}{(1+z / L)^{2}} g_{\odot}
\end{equation}
where $k=5.7$ and $L=76.8$ Mm. In this way, the plasma $\beta$ around the active region is less than 0.1 under 340 Mm. The miminum value of plasma $\beta$ is $\beta = 2.5\times 10^{-3}$. The $\rm Alfv \acute{e} n$ speed $V_{\rm A}>1000$~km~s$^{-1}$ below 190 Mm. These mimic the real corona environment better.
\subsection{Boundary Conditions}
\begin{figure}[htbp]
	\centering  
	\includegraphics[scale=0.5]{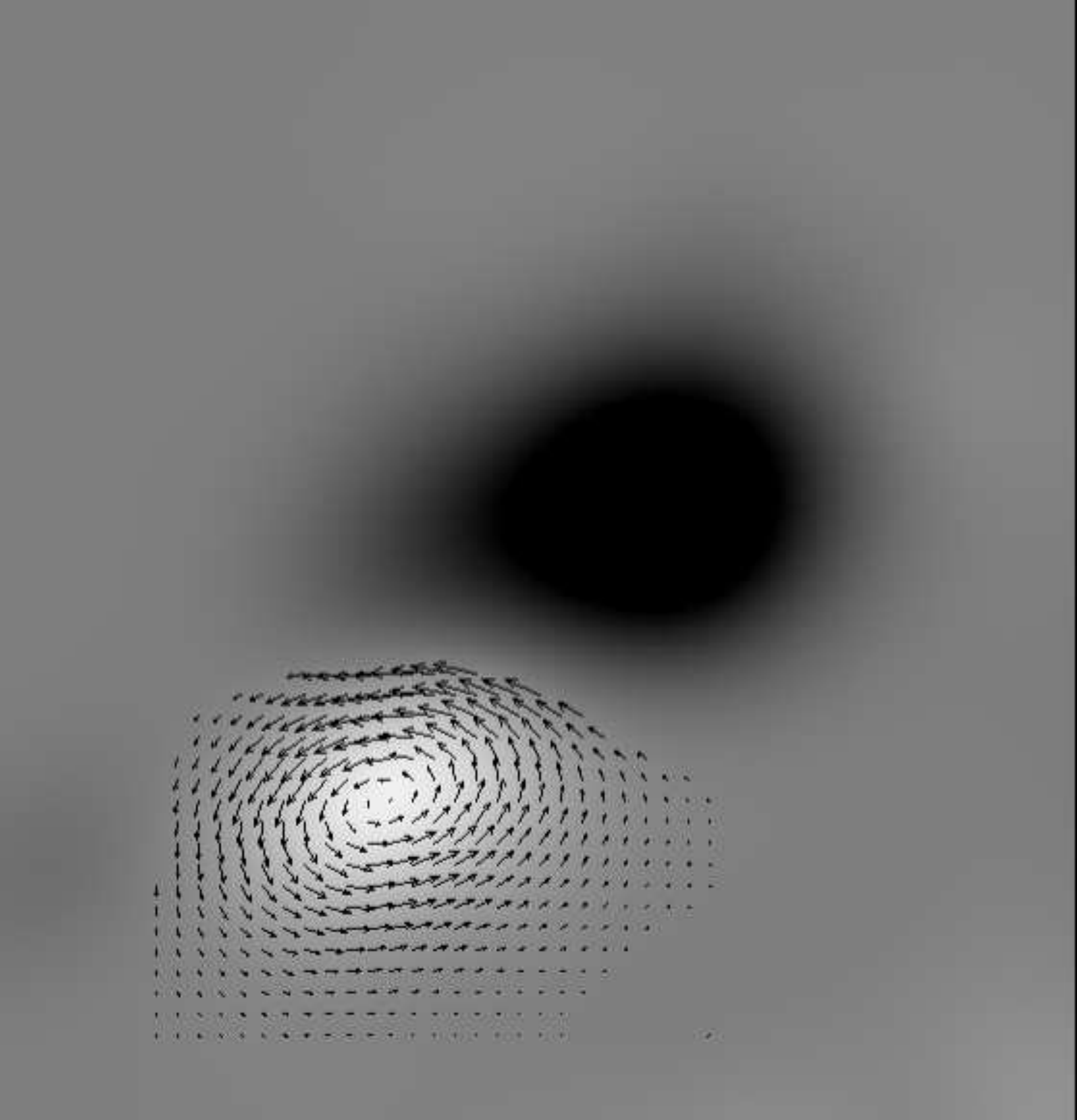}
	\caption{Velocity field at bottom boundary in our simulation.}
	\label{velocityfield}
\end{figure}

We energized the system by applying a photospheric rotational motion to the positive polarity at the bottom boundary, as shown in Figure~$\ref{velocityfield}$. To ensure that such a flow will not modify the magnetic flux distribution $B_{z}$ at the photosphere, the velocity can be specified by employing a potential function $\psi(B_{z})$ with $\mathbf v = \nabla \times (\psi \mathbf e_z) $. While the specific forms of the potential function in many previous researches (e.g., \citealt{2003ApJ...585.1073A,2010ApJ...708..314A,2013SoPh..286..453T,2021ApJ...922..108J}) made the line speed of rotation $|\mathbf v|$ vanish ($|\mathbf v|=|\nabla \psi|=0$) at PIL where $B_{z}=0$. As a result the shear flow near PIL is relative weak. To make the shear flow stronger, the velocity potential was modified as:
\begin{equation}\label{velocity potential}
\psi = v_{0}B_{z}
\end{equation}
and
\begin{equation}\label{velocity field}
v_{x}=\frac{\partial \psi\left(B_{z}\right)}{\partial y}, \quad v_{y}=-\frac{\partial \psi\left(B_{z}\right)}{\partial x}
\end{equation}
This velocity profile can reproduce the strongest shear flow $v = |\nabla \psi| = v_{0} |\nabla B_{z}|$ at PIL and the faster line speed of rotation in the north than south as observation shows \citep{2009SoPh..258..203M}. To save the computation time, $v_{0}$ is scaled such that the maximum speed is $\rm 34.1~km~s^{-1}$, which is about 60 times of the real value as $\rm 0.5~km~s^{-1}$ \citep{2009ApJ...690.1820T,2009SoPh..258..203M} but still smaller than the typical $\rm Alfv \acute{e} n$ speed by around two orders of magnitude. With such a large driving speed, the time scale of quasi-static evolution is shortened by the same times.
The photospheric motion is coupled with the magnetic field evolution by the frozen-in theorem of ideal MHD, manifested as the line-tied condition, which is important to the success of simulation. To self-consistently update the bottom magnetic field, we solve the induction equation:
\begin{equation}\label{induction equation}
\frac{\partial \mathbf{B}}{\partial t}=\nabla \times(\mathbf{v} \times \mathbf{B})+\eta_{\rm stable} \nabla_{\perp}^{2}\mathbf{B}
\end{equation}
 at the photosphere. The last term $\eta_{\rm stable} \nabla_{\perp}^{2}\mathbf{B}$ is used to maintain the numerical stability near the PIL (see also \citealt{2021FrP.....9..224J}). Here we set:
\begin{equation}\label{etastable}
 \eta_{\rm stable}=1\times 10^{-2}e^{-B^{2}_{z}}
 \end{equation}

On the side/top boundary, if we fix the plasma variables ($\rho,\mathbf{v},T$), there will be reflection. Instead, all the variables are extrapolated from the neighboring inner points using a zero gradient along the normal direction of the boundary surface. The normal component of magnetic field at side and top boundary is updated by divergence$-$free condition to avoid the accumulation of numerical error. This mimics the open boundary.
\subsection{Topology}\label{Topology}
To analyse the magnetic structure, we calculated the $Q$ factor to identify the quasi separatrix layers (QSLs) \citep{2002JGRA..107.1164T,2016ApJ...818..148L} as follows:
\begin{equation}
Q=\frac{a^{2}+b^{2}+c^{2}+d^{2}}{|a d-b c|}
\end{equation}
where
\begin{equation}
a=\frac{\partial X}{\partial x}, \quad b=\frac{\partial X}{\partial y}, \quad c=\frac{\partial Y}{\partial x}, \quad d=\frac{\partial Y}{\partial y}
\end{equation}
and $(X,Y)$ and $(x,y)$ are a pair of footpoints of the same magnetic field line. The region with large value of $Q$ (e.g., $\geq 10^5$) denotes the most possible location where reconnection will take place and is often used to be compared with the position and shape of the flare ribbons.

\section{Results}\label{Results}
\subsection{Overall Process}\label{Overall Process}
\begin{figure}[htbp]
	\centering  
	\includegraphics[scale=0.8]{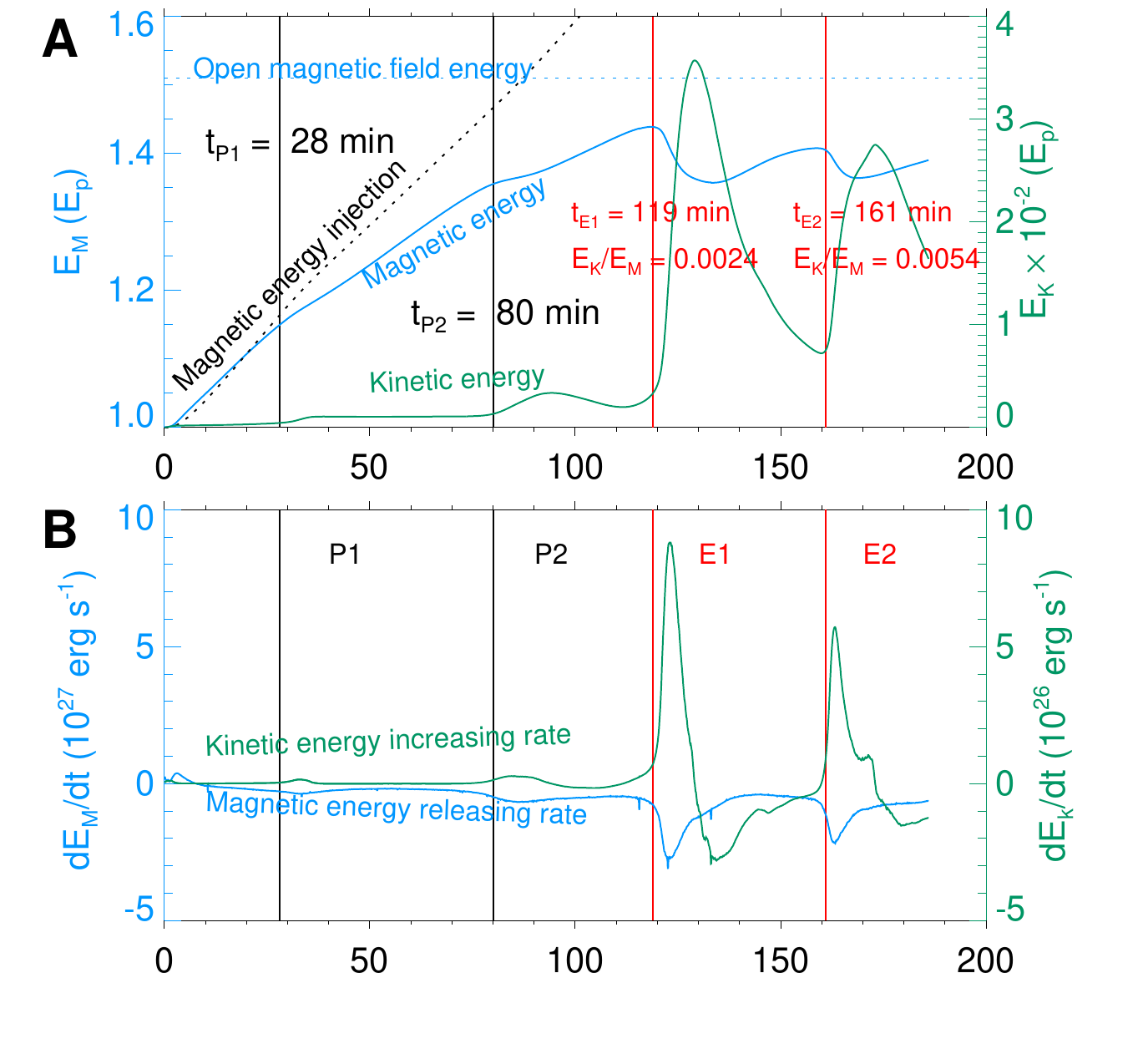}
	\caption{The energy evolution of simulation. $\mathbf{A}$): Magnetic and kinetic energy evolution during the whole process, divided into five stages. $\mathbf{B}$): Magnetic and kinetic energy changing rates corresponds to $\mathbf{A}$. In our simulation, magnetic energy of the initial potential field is $E_{\rm p}=8.3\times 10^{30}\rm erg$. This value should be multiplied by a factor of $625=\frac{1}{0.04^{2}}$ to be $\rm 5.2 \times 10^{33} erg$ which is the same order with previous results. The corresponding Supplementary Video 1 starts at $t=0$ and ends at $t=185.5$ min in simulation time, showing the evolution of $J/B$, the velocity distribution, QSLs, the magnetic field lines and the kinetic energy in the real time duration of 185.5 hours. The cadence between each figure used in Supplementary Video 1 is 210 s in the quasi-static period ($t\in [0, 112]\cup[129.5, 157.5]\cup[175, 185.5]$ min) and 21 s in the eruption period ($t\in (112, 129.5)\cup(157.6, 175)$ min) in simulation time.}
	\label{energy}
\end{figure}

Figure~\ref{energy} shows the evolution curves of the total magnetic and kinetic energies in the computational volume as well as their changing rates. The magnetic energy injection by the surface motion is also shown (the dashed line in Figure~\ref{energy}A), which is computed by time integration of the total Poynting flux at the bottom surface. As driven by the continual rotation of the sunspot for a time duration of 190~min (in which SP has rotated about 3 turns), the AR in the MHD model experiences firstly an overall increase of magnetic energy and then two eruption events with rapid release of part of the magnetic energy. The two eruptions can be identified clearly from the energy evolution, with onset time $t_{\rm E1}=119$~min for the first eruption (will be referred to as E1) and $t_{\rm E2} = 161$~min for the second eruption (E2), respectively. From the beginning to time of around $t=28$~min, the kinetic energy keeps a very low value of below $10^{-3}~E_{\rm p}$, and the magnetic energy injection curve matches well with increase of the total magnetic energy, owing to the line-tied boundary condition and the low numerical dissipation. This ideal process is followed by two small episodes of magnetic energy release that occur before E1. The first one (P1) starts at $t_{\rm P1} = 28$~min, after which the kinetic energy rises to $10^{-3}~E_{\rm p}$, and it results in a small deviation of the bottom surface energy input and the total magnetic energy accumulation. The second one (P2) occurs at $t_{\rm P2}= 80$~min, after which the kinetic energy first rises to a peak value of $3.4\times 10^{-3}~E_{\rm p}$ and then decreases slightly. The reason for these small energy release will be analyzed in the next sections.

The first major eruption (E1) begins when the magnetic energy reaches about $1.44 E_{\rm p}$ (and the sunspot has been rotated about 1.5 turns). Through this eruption, the magnetic energy decreases to about $1.36 E_{\rm p}$ ($8\times 10^{-2}~ E_{\rm p}$ free energy loss) and the kinetic energy increases impulsively to $3.6 \times 10^{-2} E_{\rm p}$. That is, about a half of the magnetic energy loss is converted to kinetic energy in 10~min. The amounts of magnetic energy released and total kinetic energy obtained, on the order of magnitude of $10^{32} $~erg, are consistent with the estimations from previous studies that used NLFFF extrapolations for the pre-flare and post-flare magnetic fields~\citep{2008ApJ...675.1637S,2010BASI...38..147R}. After the first eruption, the magnetic energy increases again while the kinetic energy drops to a low value close to that of the pre-eruption state. At $t\sim 161 $~min starts the second eruption (E2), a weaker one than the first eruption. The magnetic energy decreases from $1.41 E_{\rm p}$ to $1.36 E_{\rm p}$ ($5\times 10^{-2}~E_{\rm p} $ free energy loss). The kinetic energy increases to $2.8\times 10^{-2}E_{\rm p}$ and $56\%$ magnetic energy has been converted to kinetic energy in around 10 min. The maximum erupting speed reaches $1500$~km~s$^{-1}$ and $1100$~km~s$^{-1}$ in E1 and E2, respectively. Both eruptions drive a fast shock wave with speeds of about $500$~km~s$^{-1}$. The complex distribution of magnetic flux in this AR renders the eruptions highly asymmetrical in both north-south and west-east directions. Since the magnetic field is multiplied by a factor of 0.04, the kinetic energy should be underestimated in our simulation because if we strengthen the magnetic field used to calculate, the ratio of kinetic energy to magnetic energy will be increased.

Although it is not likely to reproduce realistically the observed flares with such simple setting of sunspot rotation, these two eruptions can still mimic approximately the observed two X-class flares on December 13th and 14th, respectively, as the first one, X3.4, is stronger than the second one of X1.5. The positive sunspot has rotated over 1.5 turns before E1, which is comparable with \cite{2009SoPh..258..203M}. Furthermore, if multiplied by a factor of 60 determined by the speeding up in our velocity-driven simulation, the quasi-static evolution time before E1 is about 5 days and time interval between E1 and E2 is about 40 hours. Both time scales are comparable with observations: 3 days of sunspot rotation before E1 and another 44-hour interval between E1 and E2. Interestingly, there will be more eruptions, produced in a homologous way, if the simulation is continued with further rotation of the sunspot, confirming that the sunspot rotation is an efficient mechanism in producing eruptions. Finally it is worthy noting that the total magnetic energy is always below the open field energy $E_{\rm open}\sim 1.51 E_{\rm p}$~\citep{1984ApJ...283..349A,1991ApJ...375L..61A,1991ApJ...380..655S} during the whole process, suggesting that eruption is efficient at keeping the magnetic energy below its upper limit, i.e., the open field energy.

\subsection{Evolution of Magnetic Field and Electric Current}\label{Magnetic Evolution}

To understand why the energies evolve in the manner as described in the last subsection, here we give a detailed study on the evolution of magnetic field, topology, and current density. First we consider the magnetic field evolution before the major eruption.

\begin{figure}[htbp]
	\centering  
	\includegraphics[scale=0.4]{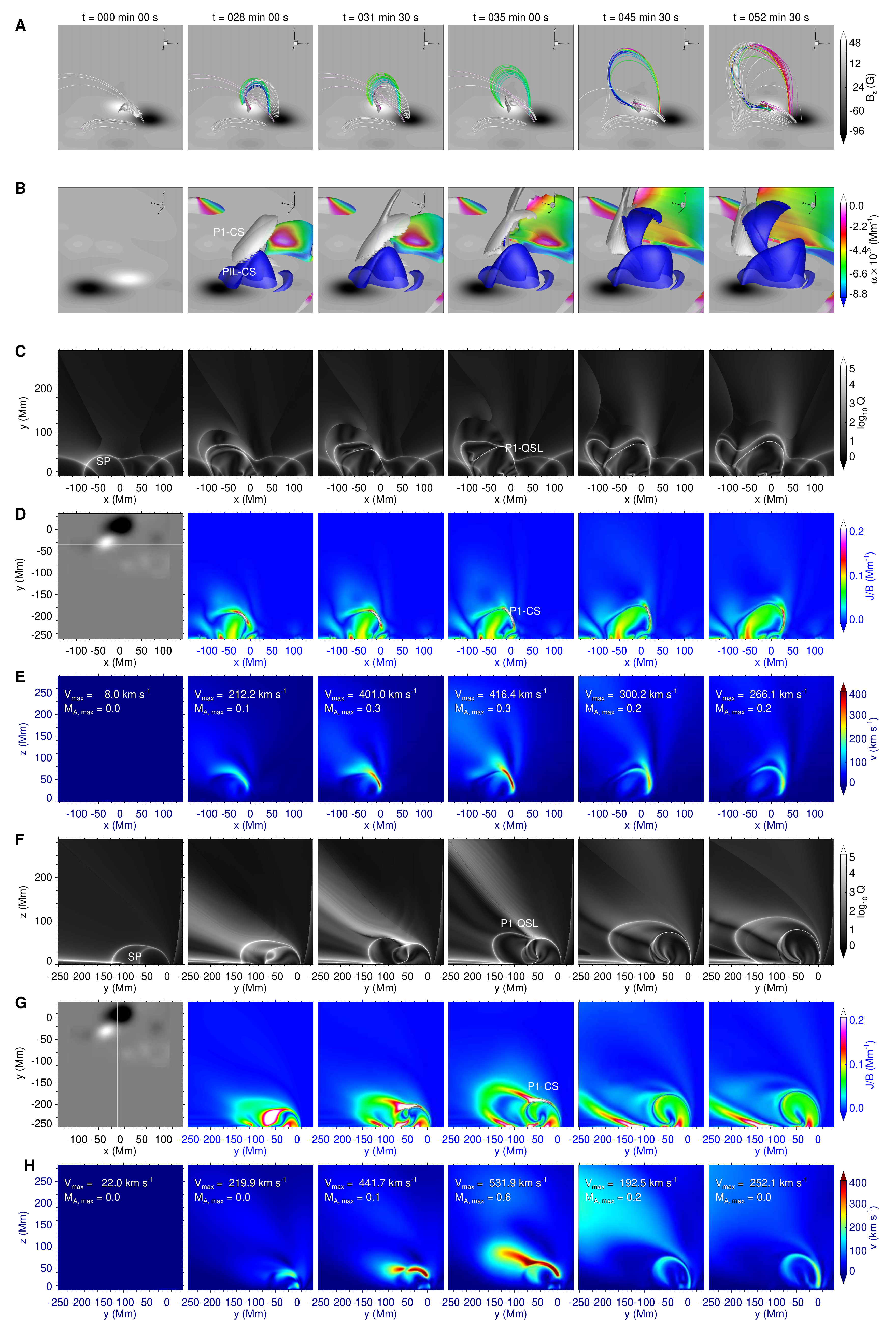}
	\caption{The magnetic evolution of P1. $\mathbf{A}$): Time series pictures of the exchanged field lines formed by reconnection in P1-CS. The color of field lines denotes the value of nonlinear force-free factor defined as $\alpha=\mathbf{J}\cdot\mathbf{B}/B^{2}$. The background shows the sunspots distribution at photosphere. $\mathbf{B}$): Evolution of iso-surface of $J/B=8.7\times 10^{-2}\rm Mm^{-1}$, which represents the current layer in different stages. $\mathbf{C}$): Slices of QSLs. $\mathbf{D}$): Slices of current layer. The initial field has no obvious current so the first column is used to label the position of slices in $\mathbf{C}$, $\mathbf{D}$ and $\mathbf{E}$. $\mathbf{E}$): Outflows at the position of current layer in $\mathbf{D}$. $\mathbf{F}$, $\mathbf{G}$ and $\mathbf{H}$): Have the same meaning with $\mathbf{C}$, $\mathbf{D}$ and $\mathbf{E}$ respectively but at the different slice as labeled by the first column of $\mathbf{G}$}
	\label{P1slice}
\end{figure}

The initial magnetic topology is shown in the first column of Figure~\ref{P1slice}C and F. It shows complex X-points around, QSLs above and at the west (right) side of SP. These initial QSLs play an important role in magnetic evolution. Before $t_{\rm P1}$, P1-QSL, i.e., the QSL above SP, was strengthened by stress between the sheared core field (which expands outward as driven by the rotation) and the surroundings. This contributed to the formation of a current layer, referred to as P1-CS at the location of P1-QSL. At the same time a current layer was also developed above the main PIL between SP and SN, for which we called the PIL-CS. It developed with rotated SP between sunspots and didn't show any sudden changes. These current layers before $t_{\rm P1}$ were not strong enough, i.e., not sufficiently thin to trigger reconnection, so the kinetic energy remained to be a very small value. The magnetic energy injection from the bottom boundary and its increase in the coronal volume matched each other well, showing the signature of quasi-static evolution in this period.

P1-CS took effect when it became strong enough after $t_{\rm P1}$. The core expansion let the P1-CS (the gray iso-surface in Figure~\ref{P1slice}B) form at the top of SP and translate to west side subsequently, with exactly the same location of P1-QSL (Figure~\ref{P1slice}C, D, F and G). Reconnection in P1-CS let WN connect to SP continuously, leading to the exchanges of SP-SN and WP-WN (as in Figure~\ref{P1slice}A). The current layer PIL-CS was still too weak before $t_{\rm P2}$ while kept developing (Figure~\ref{P1slice}B). Weak outflow ($\rm 500~km~s^{-1} $) were produced by slow reconnection in P1-CS (Figure~\ref{P1slice}E and H) which accounts for the deviation between magnetic energy injection from the bottom surface and energy accumulation between $t_{\rm P1}$ and $t_{\rm P2}$.

\begin{figure}[htbp]
	\centering  
	\includegraphics[scale=0.5]{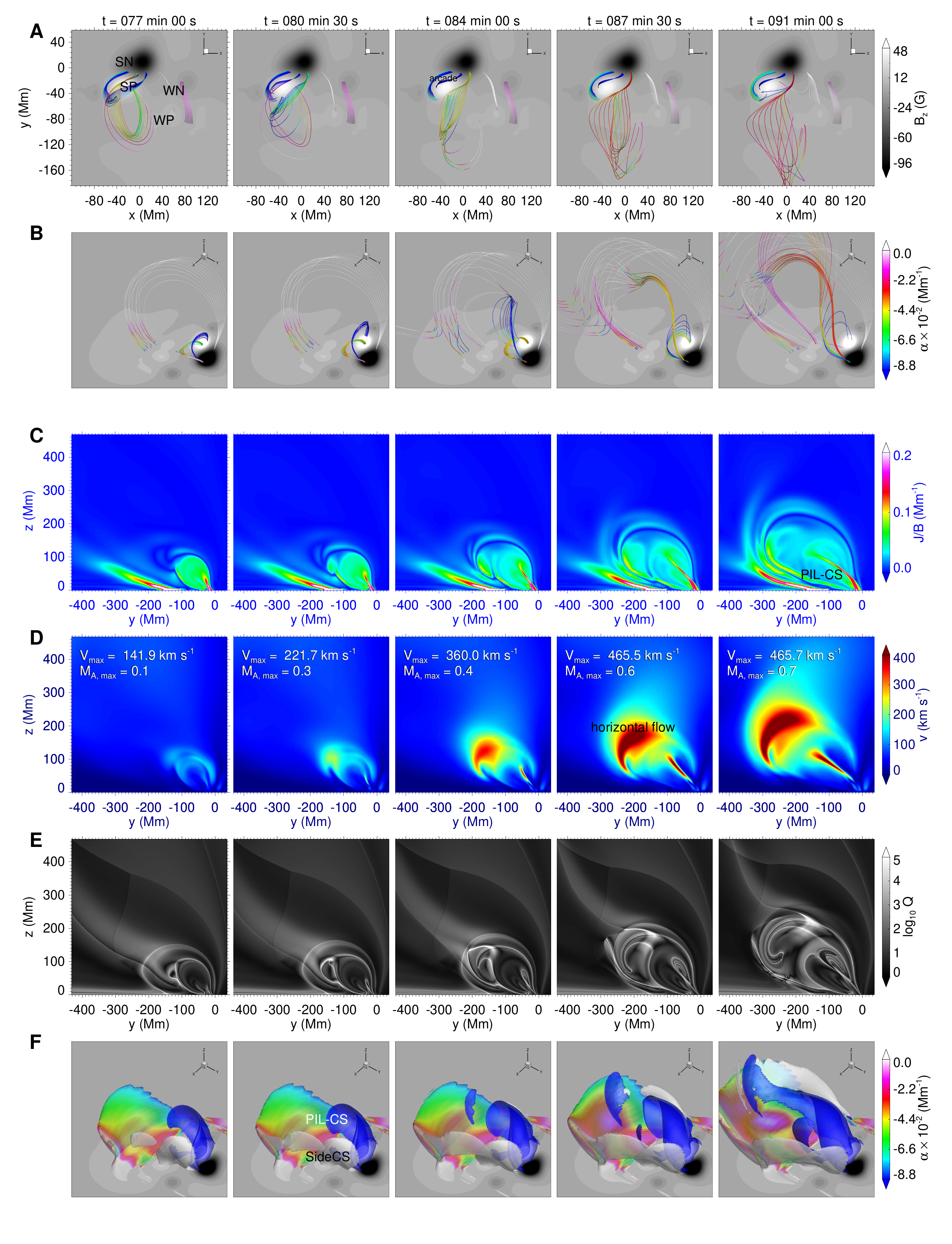}
	\caption{The trigger process of P2. Positions of all slices in $\mathbf{C}$, $\mathbf{D}$ and $\mathbf{E}$ are the same as the first column of Figure~$\ref{P1slice}$G. $\mathbf{A}$): Top view of 5 bunches of magnetic field lines with fixed negative footpoints. $\mathbf{B}$): Side view of 3D magnetic field lines. $\mathbf{C}$): Slices of current layer. $\mathbf{D}$): Outflows by slow reconnection at the position of current layer in $\mathbf{C}$. $\mathbf{E}$): Side QSLs' evolution of P2. $\mathbf{F}$):Iso-surface of $J/B=8.7\times 10^{-2}\rm Mm^{-1}$.}
	\label{P2slice}
\end{figure}

\begin{figure}[htbp]
	\centering  
	\includegraphics[scale=0.5]{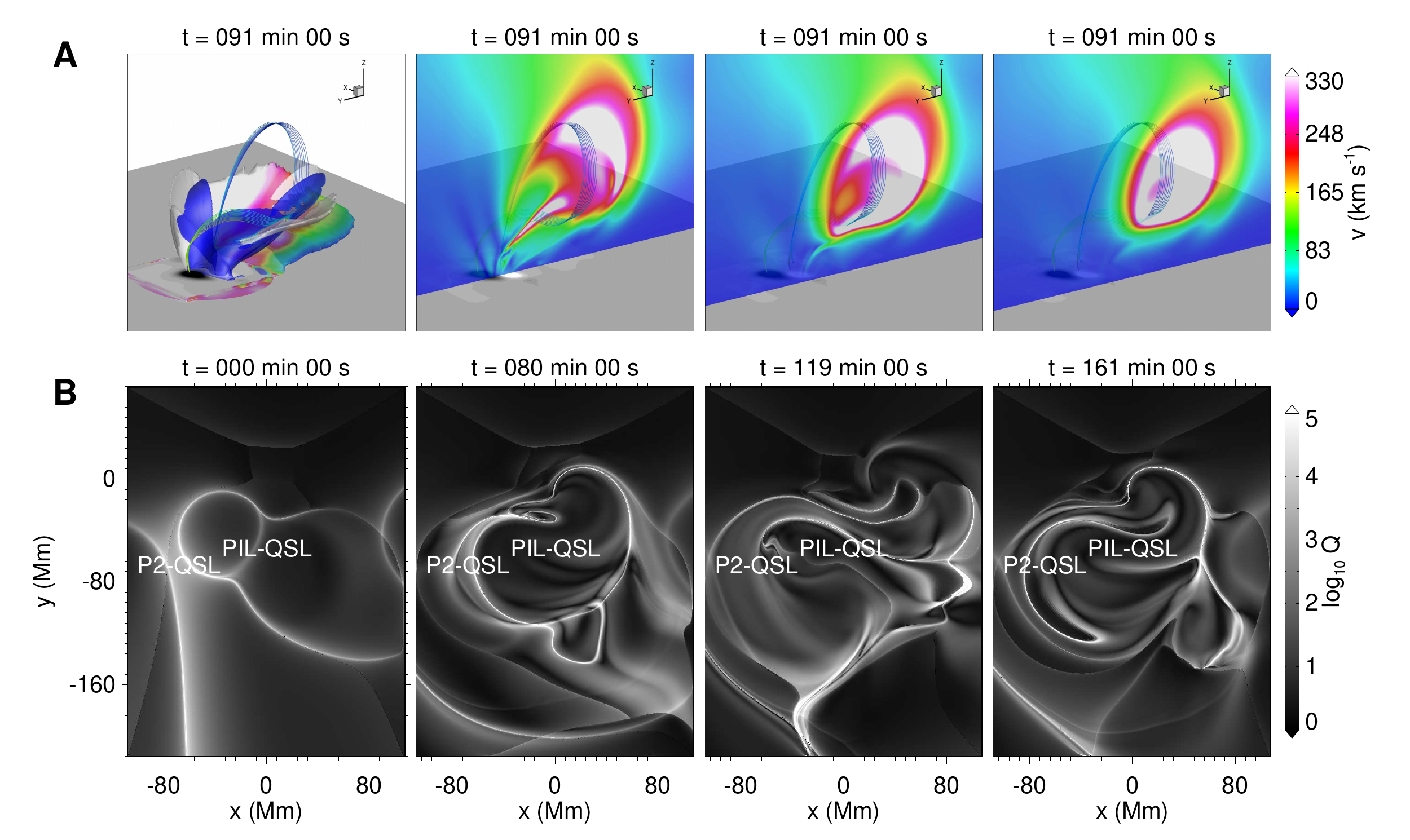}
	\caption{$\mathbf{A}$): The first column shows the MFR is located on the iso-surface of $J/B=8.7\times 10^{-2}\rm Mm^{-1}$. The 2rd to 4th columns denotes the different slices of outflow distribution. The color of iso-surface and field lines has the same meaning as previous figures. $\mathbf{B}$): QSL of $xy$-plane at the altitude which is intersect with the iso-surface in the first column of $\mathbf{A}$. Four images in order are at $t=0$, before P2, E1 and E2 respectively.}
	\label{P2flow}
\end{figure}

The third stage began after $t_{\rm P2}$. The magnetic structure is very similar with an eruption: a rising MFR, the reconnected arcades and a PIL-CS can be seen (Figure~\ref{P2slice}A, B and C). Though with these similarities, the distribution of outflow and speed show the difference (Figure~\ref{P2slice}D). The MFR is located at the PIL-CS and two parts of outflow has the same position with the intersection of the slice and MFR (Figure~\ref{P2flow}A). After we move the slice to east side, the two parts become a single one (Figure~\ref{P2flow}A). This suggests the location of reconnection was at the east side of SP. Checking the topology of magnetic structure we found that, due to the complexity of magnetic flux distribution, the initial field has a QSL (referred to as P2-QSL) at east (left) (Figure~\ref{P2flow}B). The existence of the initial P2-QSL made reconnection can take place at the location of P2-QSL before PIL-CS's width reached grid resolution. When enough magnetic field and current were transported to SP east, the second weak energy release process began with outflow speed about $\rm 500~km~s^{-1} $. Since we only rotated SP and other parts stayed nearly potential (Figure~\ref{velocityfield}), the slow outflow will be restricted by the overlying field to be the `horizontal flow' (Figure~\ref{P2slice}D), which has merely velocity in $x$ and $y$ directions. The same as in P1, the mechanism here is different from eruption: the energy conversion is resulted by the slow reconnection near the initial QSL but not fast reconnection in PIL-CS. This is the key reason why P2 is also a weak energy release episode. At the end of P2, the reconnected arcades connected to SP and SN, keeping rotating and preparing for the next eruption. Reconnection in side P1-CS existed all the time and transformed WP-WN to SP-WN (Supplementary Video 1), which let more field lines participate in the formation of PIL-CS next time. These are ready for the first major eruption (E1).

\begin{figure}[htbp]
	\centering  
	\includegraphics[scale=0.5]{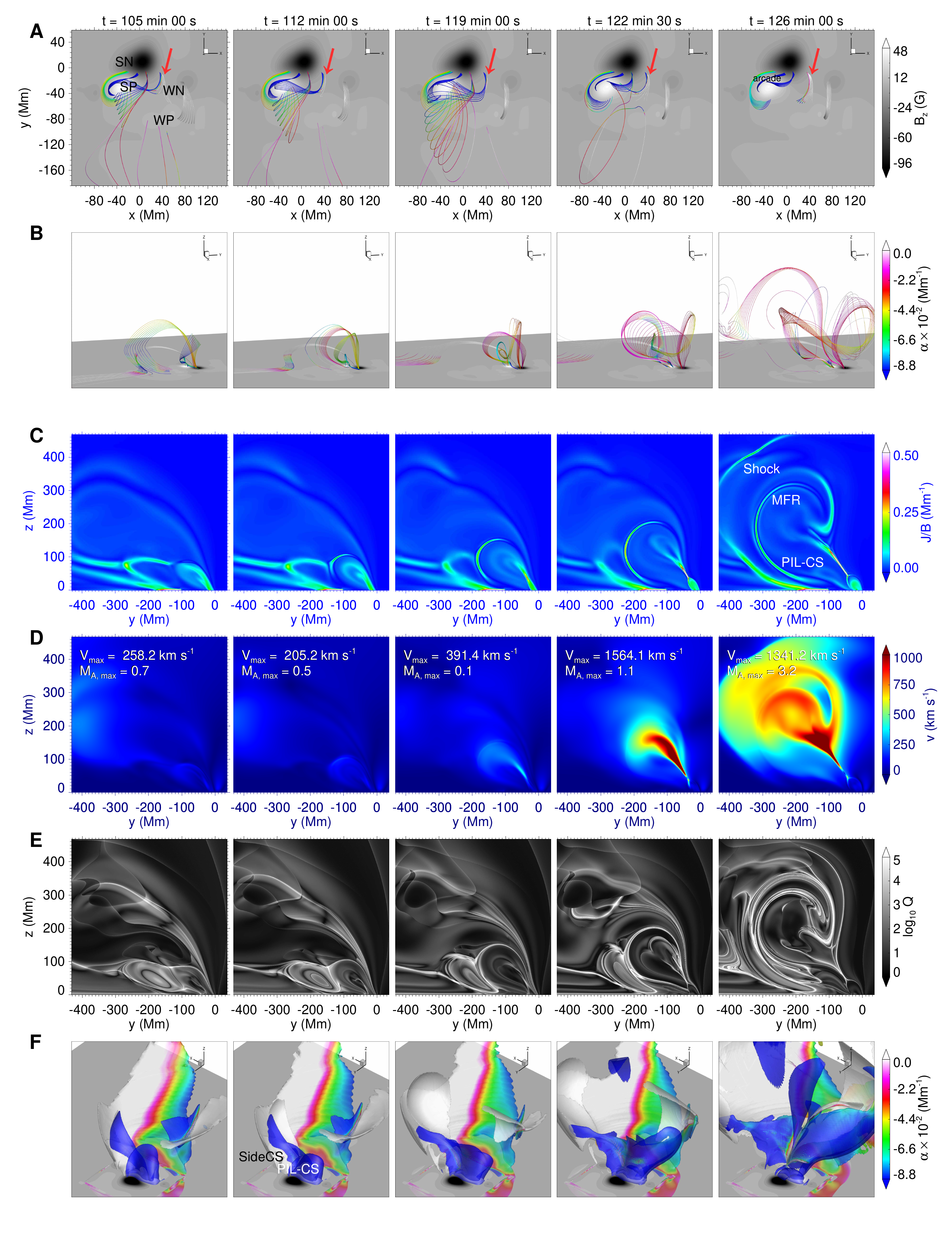}
	\caption{Magnetic evolution of E1. All settings are the same as in Figure~\ref{P2slice}. The corresponding Supplementary Video 2 starts at $t=112$ min and ends at $t=129.5$ min with a cadence of 21 s in simulation time, showing the evolution in the real time duration of 7.5 hours in E1. The magnetic field lines in video show the formation and eruption of an MFR. The slice located at $x=0$ Mm denotes the evolution of $J/B$ and the velocity distribution in video A and B, respectively.}
	\label{E1slice}
\end{figure}

\begin{figure}[htbp]
	\centering  
	\includegraphics[scale=0.5]{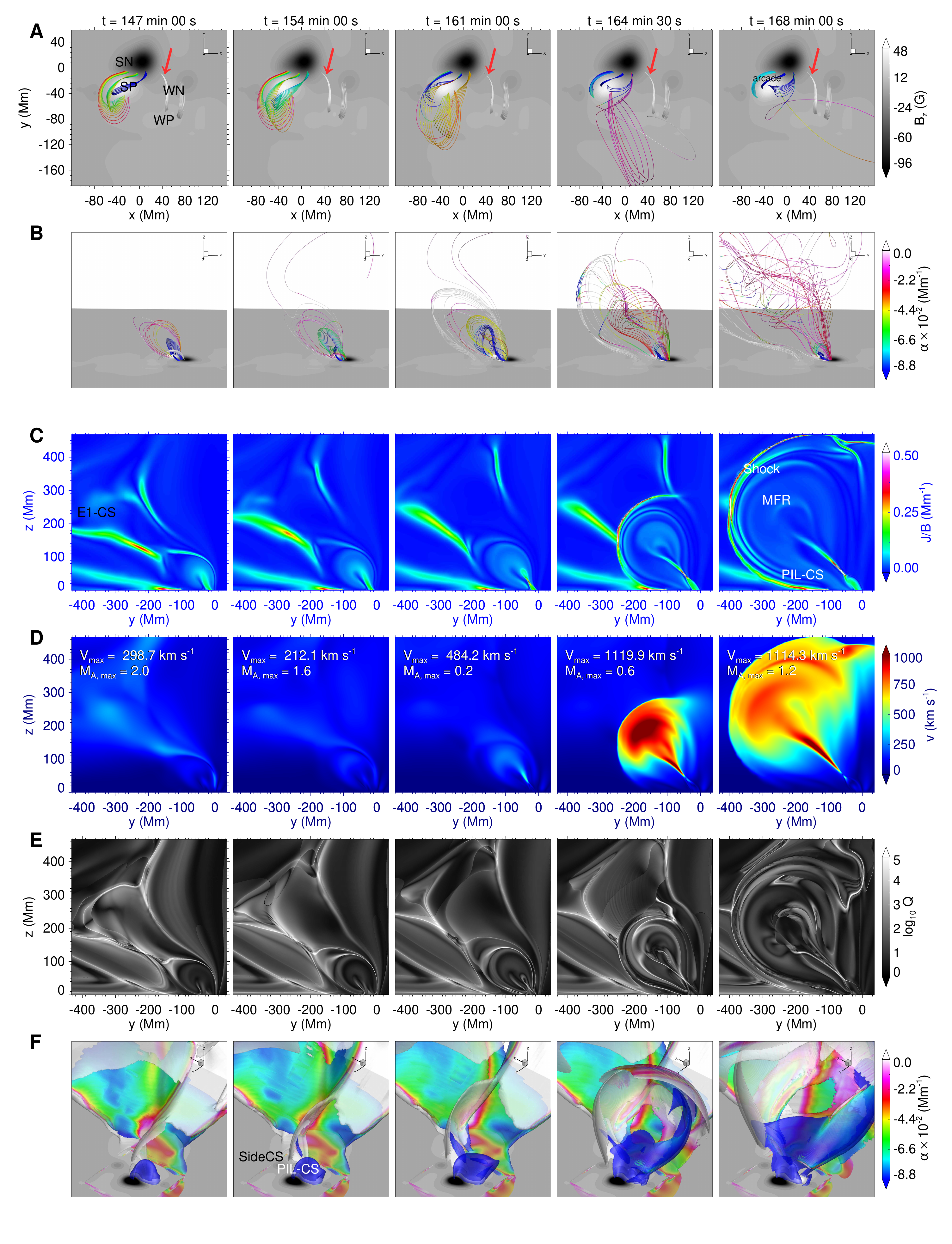}
	\caption{Magnetic evolution of E2. All settings are the same as in Figure~\ref{P2slice}. The corresponding Supplementary Video 3 starts at $t=157.5$ min and ends at $t=175$ min with a cadence of 21 s in simulation time, showing the evolution in the real time duration of 7.5 hours in E2. The magnetic field lines in video show the formation and eruption of an MFR. The slice located at $x=0$ Mm denotes the evolution of $J/B$ and the velocity distribution in video A and B, respectively.}
	\label{E2slice}
\end{figure}

During the periods as described above, converging motion towards PIL induced by rotation kept thinning the CS between SP and SN (PIL-CS) with a speed as the same order of rotation, i.e., 2 orders of magnitude lower than local $V_{\rm A}$ ($\rm Alfv \acute{e} n$ speed), thus representing the `quasi-static evolution'. Owing to the very low magnetic diffusion in our code, we can get a very thin CS even with such a low speed. Otherwise, a larger magnetic diffusion will widen the CS against the converge motion, as pointed out in \cite{2021NatAs...5.1126J}. The fourth bunch of field lines (labeled by the red arrow in Figure~\ref{E1slice}A) became SP-WN (Figure~\ref{E1slice}A) and took effect to form a stronger PIL-CS by rotational post-P2 arcades. The trigger PIL-CS grow up from the bottom of simulation box by continuous rotation (Figure~\ref{E1slice}C) until the thickness of PIL-CS reached 2-3 grid resolution. Then numerical diffusion became non-negligible and triggered the fast reconnection and the fourth stage, namely, the major eruption (E1) began.  The PIL-CS also extended to WN (Figure~\ref{E1slice}F), which corresponds to the longer flare ribbon on December 13th 2006. Reconnection in PIL-CS formed an MFR during the eruption (Figure~\ref{E1slice}B). The plasma outflow originated from the PIL-CS with a speed reaching up to $\rm 1500~km~s^{-1} $ (Figure~\ref{E1slice}D) and impulsively drove MFR to erupt (as shown in Supplementary Video 2A). Meanwhile, the PIL-CS became longer in the vertical direction, thinner and stronger (Figure~\ref{E1slice}C and Supplementary Video 2B), with more flux involved into reconnection, which provides the energy required for this eruption. With such a high speed, this eruption was strong enough to remove the restriction of overlying field (which also occurred in E2) and no `horizontal flow' can be seen in Figure~\ref{E1slice}D and \ref{E2slice}D. At the end of E1, the reconnected arcades SP-SN restored and kept rotating as before (Figure~\ref{E1slice}A). While SP-WN returned to its origin WP-WN (labeled by the red arrow in Figure~\ref{E1slice}A) and was out of control of rotation.

When the arcades SP-SN formed after E1 was sheared enough again, the fifth stage began (after $t_{\rm E2}$). The same as in E1, the PIL-QSL along with PIL-CS grew up again from bottom near PIL. Reconnection in PIL-CS formed an MFR (Figure~\ref{E2slice}B), which was lift up by the outflow (with the speed of $\rm 1100~km~s^{-1}$) initiated from PIL-CS  (Figure~\ref{E2slice}D and Supplementary Video 3A). The side P1-CS always existed and transformed the field line connection of WP-WN to that of SP-WN, while the time duration of side reconnection before $t_{\rm E2}$ was not so long as that before $t_{\rm E1}$. As a result, less magnetic flux was involved in the formation of PIL-CS (the field lines labeled by red arrow in Figure~\ref{E2slice}A remain WP-WN), which made the eruption CS weaker than E1 (Figure~\ref{E2slice}C) and shorter in the vertical direction (Supplementary Video 3B). The flare ribbon and PIL-CS (Figure~\ref{E2slice}F) were shorter also in the horizontal direction. This naturally leads to the fact that the magnetic energy release in E2 is less than that in E1. We note that when E2 began, the current sheet of E1 didn't disappear (Figure~\ref{E2slice}C), and in a short interval, the latter eruption (E2) caught up with the former one (E1), making the shock in E2 clearer. After the eruption, the post-flare arcades should restore to the pre-flare configuration again, and if with further rotation of the sunspot, it will lead to the third eruption which is beyond the scope of this event research. We stopped the simulation at $t\sim190$ min, showing the whole process of magnetic evolution of 2 eruptions and the reasons for such changes.

\subsection{Comparison with Observations}\label{observations}
To show the credibility of our simulation, our results are compared with the observed X-ray and $\rm H\alpha$ features, time scale, rotation angle, magnetic energy release and relative strength of the eruptions.

\begin{figure}[htbp]
	\centering  
	\includegraphics[scale=0.6]{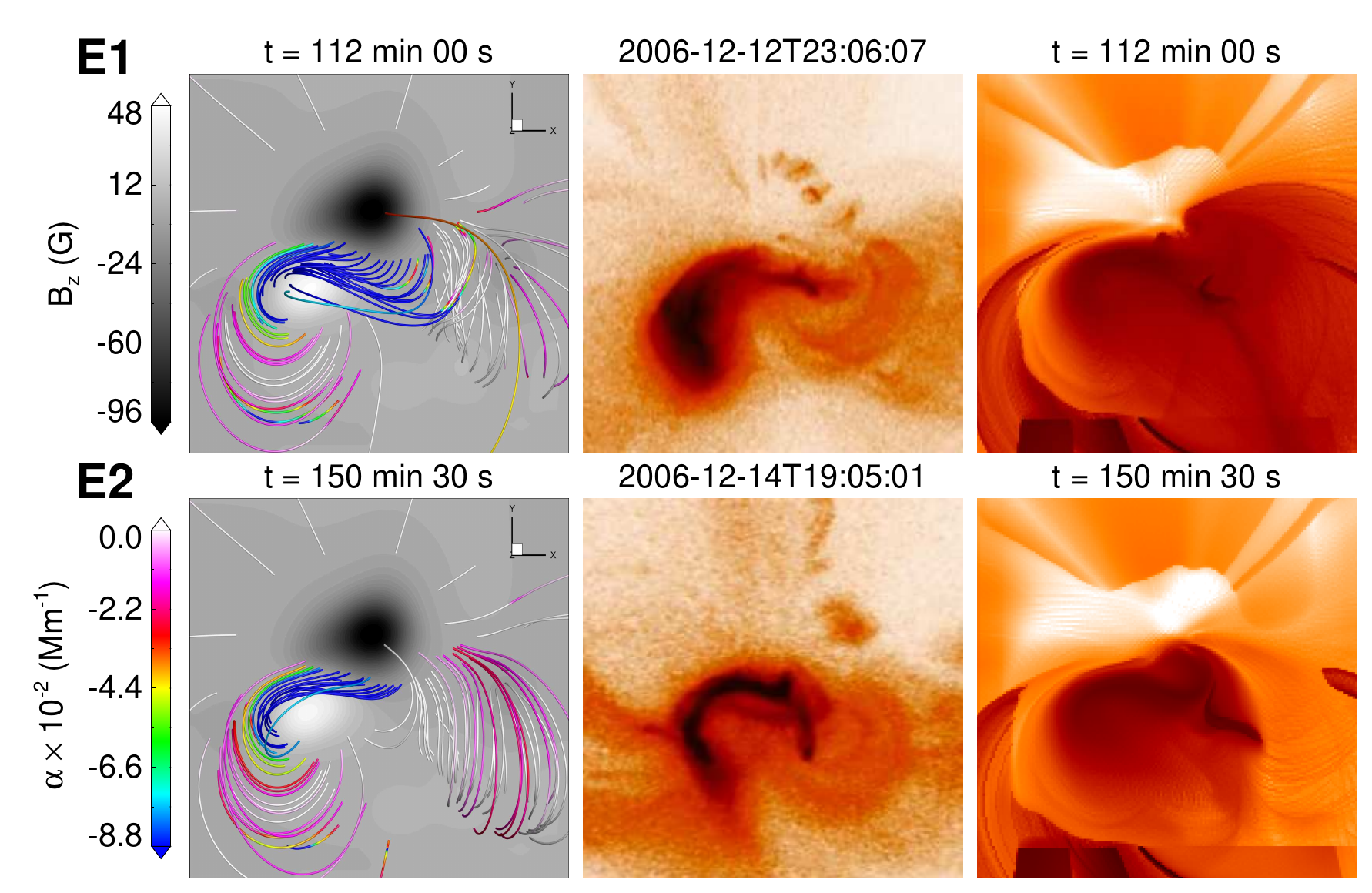}
	\caption{Comparison of simulation results with observation. $\mathbf{E1}$): In order are field lines, XRT sigmoid and the synthetic image of coronal emission from current density respectively before the first eruption. $\mathbf{E2}$): Same as E1 but before the second eruption.}
	\label{obslice}
\end{figure}

In general, QSLs denotes the location where reconnection is most likely to take place and their footpoints at the bottom surface represent the position of flare ribbons \citep{2002JGRA..107.1164T}. Figure~\ref{QSLflare} shows the comparison of the bottom QSLs for the two simulated eruptions with the flare ribbons as observed for the two flares. During both eruptions, the QSLs are overall consistent in shape and position with observed flare ribbons: QSL-N (corresponding to the negative flare ribbon) was located near the main PIL between two sunspots. QSL-P (corresponding to the positive ribbon) was initially at the east (left) of the positive sunspot and then shrank into a quasi-circular shape to the west (right). Both are comparable with the evolution of flare ribbons, especially the QSL-N, which extended longer in E1 than E2 (Figure~\ref{QSLflare}). This was formed by the side reconnection in P1-CS, which transformed field connection of WP-WN to SP-WN as described in Section \ref{Magnetic Evolution} and as a result the west field lines were involved in the eruption. Sigmoids in soft X-ray images (Figure~\ref{obslice}) before both flares were located at the main PIL and bent towards the positive sunspot by sunspot rotation. These observed features are comparable with the synthetic images of coronal emission from current density \citep{2016ApJ...828...62J} and simulated magnetic structure (Figure~\ref{obslice}).

Quantitatively, the simulation can also yield consistence in timing and magnetic energy release as mentioned in Section~\ref{Overall Process}. There were 5 days rotation before E1 and another 40-hour interval between E1 and E2. Both time scales are comparable with actual evolution time: 3 days rotation before E1 and 44 hours time interval between eruptions on December 13th and December 14th. The positive sunspot has rotated over 1.5 turns before the first eruption in our simulation, which is consistent with the total rotational angle of $540^{\circ}$ as derived in \cite{2009SoPh..258..203M}. The magnetic energy release in E1 in our simulation is $\Delta E_{\rm mag}=3.6\times 10^{32}~\rm erg$, which is very close to the values derived with other methods in previous researches of $\Delta E_{\rm mag}\sim 3\times 10^{32}~\rm erg$ \citep{2008ApJ...675.1637S,2010BASI...38..147R}. From observation, the CME on December 13th (1780 km~s$^{-1}$) is faster than on December 14th (1042 km~s$^{-1}$), which is consistent with our simulation.

\begin{figure}[htbp]
	\centering  
	\includegraphics[scale=0.6]{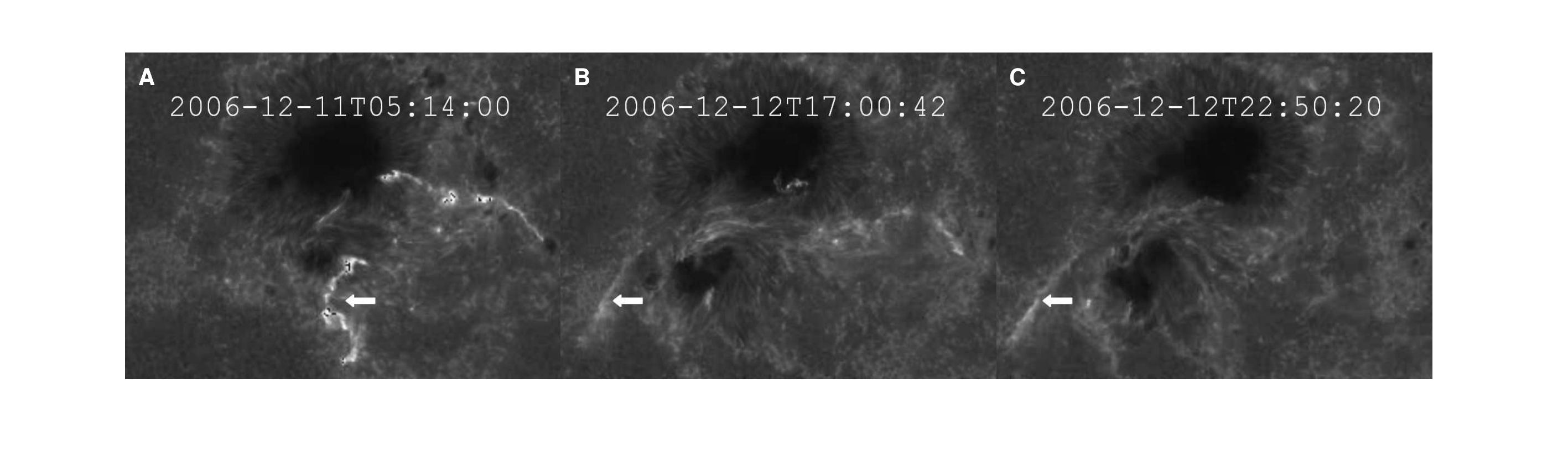}
	\caption{H$\alpha$ ribbons images taken from SOT/BFI. $\mathbf{A}$): The H$\alpha$ brightening at the similar location of P1-CS. $\mathbf{B}$): The H$\alpha$ brightening at the similar location of P2-QSL. $\mathbf{C}$): Same as B but at a different time.}
	\label{ob_p1p2}
\end{figure}

The H$\alpha$ figures also show some observational evidence corresponding to P1 and P2 episodes as labeled by the white arrow in Figure~\ref{ob_p1p2}. The H$\alpha$ brightening has the similar location of P1-CS in Figure~\ref{ob_p1p2} A and P2-QSL in Figure~\ref{ob_p1p2}B and C respectively. This indicates the slow reconnection there before the major eruption as described in Section \ref{Magnetic Evolution}. These results enhance the credibility of our simulation.

It should be noted that our simulation simplified the photospheric motions in many aspects, which could affect the results. We did not include the flux emergence process of the rotating sunspot, its shearing motion (from west to east) with respect to the leading sunspot SP, and the colliding motion between the two main sunspots \citep{2008ApJ...687..658W}. For example, if we move the positive sunspot from west to east, the QSL-N in E1 may be longer since when the positive sunspot is located further east, it will connect to WN with a stronger sheared configuration. Larger computational domain is also helpful to obtain a longer QSL: once the MFR reaches the top or lateral boundaries, the closed field lines will be taken as the open field and can't be shown by Q factor calculation. These adjustment has potential to get a higher degree of consistence between simulated QSLs and observed flare ribbons. Also the converge motions (i.e., the collision of the two sunspots) will shorten the evolution time since it will enhance the building up of the PIL-CS and enhance the amount of the magnetic energy release by strengthening the magnetic gradient near PIL \citep{2022A&A...658A.174B}. Though more complex motions and settings may reproduce the flares more realistically, our result shows the key role played by sunspot rotation in leading to the eruptions and can shed light on the onset mechanism of this homologous event.

\subsection{Eruption Initiation Mechanism}\label{trigger}
There are two types of CS in our simulation: the formation of CSs in P1 and P2 which are responsible for slow reconnection depends on the initial topology while PIL-CS formed by continuous shear near PIL which accounts for the main energy release in E1 and E2. As the sunspot rotation brought field lines together, the magnetic field expanded slowly in P1 and was translated to P2-QSL in P2. This leaded to the squeeze between core field and the surroundings. Then a squeezed QSL formed at top (Figure~\ref{P1slice}G) and the east (left) side of the positive sunspot (Figure~\ref{P2flow}). Slow reconnection here changed the magnetic topology without eruptions. During the same period, converging motion induced by rotational flow made PIL-CS stronger and thinner. When the CS's thickness reached down to the grids width, magnetic gradient near CS will be strong enough to let the diffusion kick in. This mimicked essentially the non-uniform magnetic diffusivity as required in the Pescheck-type reconnection \citep{1994ApJ...436L.197Y}: the resistivity depends sensitively on the local current density, and finally leaded to fast reconnection and eruption.

\begin{figure}[htbp]
	\centering  
	\includegraphics[scale=0.8]{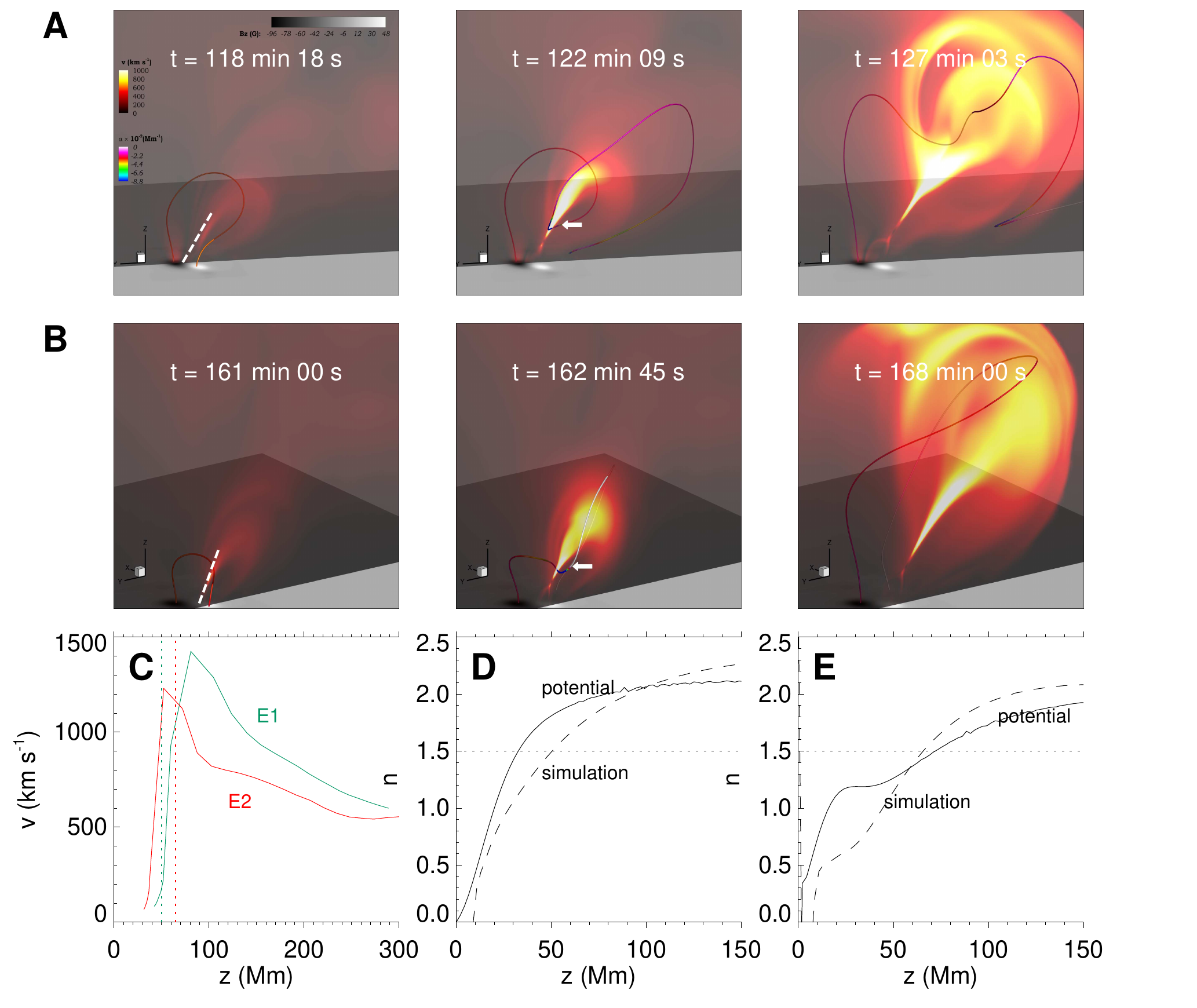}
	\caption{$\mathbf{A}$): 3 snapshots of the evolution of reconnected field lines in E1. The vertical slice at $x=0$ (at the middle of MFR) is shown with the contour of velocity. All physical variables are displayed by the colorbar in the first panel. $\mathbf{B}$): Same as A but the vertical slice is located at $x=-34.5$ Mm. $\mathbf{C}$): The solid lines denote the variation of the speed with respect to height $z$ of the point which is the intersection of the reconnected field lines and the vertical slice in A (E1) and B (E2), respectively. The green (red) dashed line denotes the critical height of torus instability in E1 (E2). $\mathbf{D}$): Variation of decay index $n$ with respect to height $z$ along the white dashed line on the slice in the first panel of A. The dashed line in D denotes the decay index $n$ of simulated field at $t\sim $118 min and the solid line denotes the result of the corresponding potential field at the same time. $\mathbf{E}$): Same as D but $n$ is calculated along the white dashed line in the first panel of B at $t\sim$161 min.}
	\label{MFRv}
\end{figure}

Figure~\ref{MFRv}C shows the temporal evolution of velocity at approximately the middle point of the field lines as shown in Figure~\ref{MFRv}A and B, respectively. These field lines are used to illustrate the dynamics of the field that experienced reconnection and became part of the MFR subsequently in the two eruptions, E1 and E2. Once the reconnection took place, the coronal plasma as frozen with the field lines was accelerated impulsively from a few 10 $\rm km~s^{-1}$ to over 1000 $\rm km~s^{-1}$. This acceleration was accomplished by the strong slingshot effect of the upward concave magnetic field lines as labeled by the white arrow in the middle panel of Figure~\ref{MFRv}A and B. Shortly after the impulsive acceleration, the upward tension force changed sign to a downward one since the magnetic field lines relaxed quickly from upward to downward concave shape. As a result, the field lines experienced deceleration from above 1000 $\rm km~s^{-1}$ to around 600 $\rm km~s^{-1}$, which is consistent with the MFR acceleration process described in \cite{2021NatAs...5.1126J}, suggesting the magnetic reconnection played the key role in initiating the two eruptions.

We also estimated the possible role played by torus instability in driving the eruptions of E1 and E2. To do this, we need to calculate the decay index $n$ of the strapping field (often approximated by the potential field model) overlying the erupting MFRs. Since the potential field is not always a good approximation of the strapping field (especially when the overlying field is substantially sheared), we also calculated the decay index of our simulated field for comparison in Figure~\ref{MFRv}D and E. The decay index was derived along the white dashed line in Figure~\ref{MFRv}A and B, which denotes the eruption direction following the method proposed by \cite{2019ApJ...884...73D}. The critical height of simulated field (above which $n>1.5$) is located at 50 Mm in E1 and above 60 Mm in E2. The reconnection point (labeled by the white arrow in the middle panel of Figure~\ref{MFRv}A) is located at the height of 50-60 Mm in E1, which indicates the MFR axis entered the unstable region. Therefore, when the MFRs in E1 was formed, the torus instability was possible to be triggered to drive the eruption in addition to the reconnection. While the MFR axis in E2 is located below 60 Mm and the torus instability had little chance to take effect. This may be an additional reason why E1 is stronger than E2, as the overlying field of E1 decays faster with height than that in E2. It is also worthy noting that, PIL-CSs were formed before the onset of both eruptions, or in other words, they were all formed in a quasi-static way before MFR exists. Then an MFR was formed synchronously with the reconnection and acceleration in PIL-CSs (Supplementary Video 2A and 3A). The acceleration process of the erupting MFR was accomplished under the critical height of torus instability in E2 while above the critical height in E1 as shown in Figure~\ref{MFRv}C. Furthermore, the MFRs experienced a deceleration process after the impulsive acceleration phase, and this deceleration occurs even in the torus unstable region of two eruptions, which clearly indicated the torus instability was not the main factor controlling the dynamics of the MFRs. Therefore, though torus instability had the potential to be triggered and helped the acceleration in E1, magnetic reconnection was the main initiation mechanism of both eruptions.

The P1-CS formed at the top of the positive sunspot initially and the four polarities: SP, SN, WP and WN constituted a quadrupolar topology. One may compare this situation with the breakout model: P1-CS corresponds to the breakout CS which opens the overlying field of the eruptive core in the quadrupolar configuration. However, our case is unlike the breakout model in which the reconnection at the breakout CS plays the key role in triggering eruption. In our simulation, the main consequence of slow reconnection in P1-CS is changing the magnetic connectivity, which can make E1 stronger, but it is not required to trigger the eruption. Continuous sunspot rotation can initiate the eruption alone from sheared PIL-CS. This clearly suggests that the mechanism as demonstrated here is a fundamental one, which is consistent with that shown in \cite{2021NatAs...5.1126J} and \cite{2022ApJ...925L...7B}.
\section{Conclusions and Discussions}\label{Conclusion}
In this paper, using our velocity-driven DARE-MHD model, we have simulated the two eruptions of NOAA AR 10930 on December 13th 2006 and December 14th 2006 continuously. Our simulation started from a potential field obtained by the observed magnetogram and with a simple rotation flow applied to one of the main magnetic polarities at the bottom surface to mimic the sunspot rotation. Owing to the complex distribution of the magnetic flux, there were two slow reconnection processes in P1 and P2 before the first major eruption, which helped building the special magnetic topology. When sunspot rotated over 1.5 turns, the most strong CS formed near the main PIL. Fast reconnection in PIL-CS formed an MFR and the reconnection outflow ejected the coronal plasma violently. The PIL-CS was stretched to be longer, stronger and thinner, and continuous reconnection released the energy required by E1. After this eruption, the post-flare arcades of E1 were further stressed by the rotating sunspot with about another half turn, during which the PIL-CS forms again and  then the second major eruption began, which is very similar to the homologous eruption mechanism as shown in \cite{2022ApJ...925L...7B}. The PIL-CS between sunspots was developed from bottom and reconnection sets in to trigger the eruption when the width was comparable with grid resolution like in E1. Though E1 and E2 had the same mechanism, there were less magnetic flux participated in E2, which made the CS and also the eruption in E2 weaker than in E1. Two eruptions have reasonable consistency with observations in relative strength, magnetic energy loss, sunspot rotation angle in the pre-flare duration, observed X-ray and $\rm H\alpha$ features, as well as eruption time interval.

Our simulation offers a scenario different from many previous studies. For example, different from \cite{2014Natur.514..465A}, in which they drove an NLFFF extrapolated for about 6 hours before the eruption to erupt by using three different types of photospheric boundary conditions, we started the simulation from the potential field. Moreover, the continuous energy accumulation and release process has been produced in a more self-consistent way, by applying a more realistic condition, i.e., the rotation of the positive sunspot at the bottom boundary. The key magnetic structure in favor of initiating the eruption can form by sunspot rotation directly. In addition, a pre-eruption MFR that emerged through the photosphere as described in \cite{2011ApJ...740...68F} is not necessary in our simulation for triggering the eruption. The sunspot rotation can form an MFR by fast reconnection in the pre-formed PIL-CS during E1 and E2, but not before. Furthermore, most researches of AR 10930 are only focused on the X3.4 flare on December 13th and few of them have studied the relationship of the two flares (the other X1.5 flare on December 14th). Our results show the two events can be triggered in the same way by fast reconnection in the CS formed in a recurrent manner by sunspot rotation as described in Section~\ref{Results}. Since sunspot rotation is a persistent motion for days, our result suggests an efficient way of continuous energy injection, which can reproduce the homologous eruption in AR 10930.

The importance of sunspot rotation has also been taken into consideration in some previous researches while the corresponding numerical models were established in different ways. To investigate the effect of sunspot rotation in AR 10898, \cite{2013SoPh..286..453T} rotated a envelop field of a pre-existing MFR. As the the envelop field expanded progressively, the MFR became unstable and triggered to erupt by torus instability. \cite{2021ApJ...922..108J} rotated the reconstructed potential field of AR 12665 along with flux emergence at the PIL. As a consequence, a sigmoidal structure formed with an overlying MFR created and rose to erupt like a CME. Compared with these rotation-driven models, our model is much simpler with only sunspot rotation and a potential field as the initial state, and the MFRs in our simulation could form spontaneously at the onset time of reconnection in the pre-formed PIL-CS and erupted as a CME. Both of the formation of MFR and PIL-CS were not related to the flux emergence. The quasi-static evolution and impulsive eruption process can be obtained solely by the rotation of the initial potential field.

Owing to the simple settings of our simulation in many aspects, more realistic consideration should be taken in future improvements of the model for reproducing the eruptions. For example, the smoothing of magnetogram have weakened the magnetic gradient near the main PIL. As a consequence the eruption strength will decrease, because according to \cite{2022A&A...658A.174B}, the eruption strength is highly correlated with the magnetic gradient of the main PIL. A more realistic velocity field at photosphere, including rotational, shearing and converge motions, derived from observation could be applied as the boundary condition to get a more self-consistent and realistic evolution (which has been shown in \citealt{2021FrP.....9..224J} and \citealt{pub.1147837699}). Another key point we have not considered in the current model is the flux emergence process, for which the normal velocity (i.e. velocity in $z$-direction) at photosphere ought to be used to mimic the emergence process of sunspot.

To summarize, the whole process from potential field to eruption has been reproduced, showing the full MHD evolution of slow energy accumulation to fast release. Our simulation reveals the importance of sunspot rotation and magnetic reconnection in eruption initiation mechanism. The homologous eruptions as driven by persistent photospheric motion and initiated by the fundamental mechanism \citep{2021NatAs...5.1126J} may be common in solar ARs. Future works will be carried out with the aforementioned improvements for more realistic modeling of solar eruptions that can be potentially applied to the space weather forecast.
~\\

This work is jointly supported by National Natural Science Foundation of China (NSFC 41731067, 42030204, 42174200), the Fundamental Research Funds for the Central Universities (Grant No. HIT.OCEF.2021033), Shenzhen Science and Technology Program (Grant No. RCJC20210609104422048 and JCYJ20190806142609035). The computational work was carried out on TianHe-1(A), National Supercomputer Center in Tianjin, China.


\begin{thebibliography}{1}
\expandafter\ifx\csname natexlab\endcsname\relax\def\natexlab#1{#1}\fi
\providecommand{\url}[1]{\href{#1}{#1}}
\providecommand{\dodoi}[1]{doi:~\href{http://doi.org/#1}{\nolinkurl{#1}}}
\providecommand{\doeprint}[1]{\href{http://ascl.net/#1}{\nolinkurl{http://ascl.net/#1}}}
\providecommand{\doarXiv}[1]{\href{https://arxiv.org/abs/#1}{\nolinkurl{https://arxiv.org/abs/#1}}}

\bibitem[{{Aly}(1984)}]{1984ApJ...283..349A}
{Aly}, J.~J. 1984, \apj, 283, 349, \dodoi{10.1086/162313}

\bibitem[{{Aly}(1991)}]{1991ApJ...375L..61A}
---. 1991, \apjl, 375, L61, \dodoi{10.1086/186088}

\bibitem[{{Amari} {et~al.}(2014){Amari}, {Canou}, \&
  {Aly}}]{2014Natur.514..465A}
{Amari}, T., {Canou}, A., \& {Aly}, J.-J. 2014, \nat, 514, 465,
  \dodoi{10.1038/nature13815}

\bibitem[{{Amari} {et~al.}(2003){Amari}, {Luciani}, {Aly}, {Mikic}, \&
  {Linker}}]{2003ApJ...585.1073A}
{Amari}, T., {Luciani}, J.~F., {Aly}, J.~J., {Mikic}, Z., \& {Linker}, J. 2003,
  \apj, 585, 1073, \dodoi{10.1086/345501}

\bibitem[{{Antiochos} {et~al.}(1999){Antiochos}, {DeVore}, \&
  {Klimchuk}}]{1999ApJ...510..485A}
{Antiochos}, S.~K., {DeVore}, C.~R., \& {Klimchuk}, J.~A. 1999, \apj, 510, 485,
  \dodoi{10.1086/306563}

\bibitem[{{Aulanier} {et~al.}(2010){Aulanier}, {T{\"o}r{\"o}k}, {D{\'e}moulin},
  \& {DeLuca}}]{2010ApJ...708..314A}
{Aulanier}, G., {T{\"o}r{\"o}k}, T., {D{\'e}moulin}, P., \& {DeLuca}, E.~E.
  2010, \apj, 708, 314, \dodoi{10.1088/0004-637X/708/1/314}

\bibitem[{{Bamba} {et~al.}(2013){Bamba}, {Kusano}, {Yamamoto}, \&
  {Okamoto}}]{2013ApJ...778...48B}
{Bamba}, Y., {Kusano}, K., {Yamamoto}, T.~T., \& {Okamoto}, T.~J. 2013, \apj,
  778, 48, \dodoi{10.1088/0004-637X/778/1/48}

\bibitem[{{Bateman}(1978)}]{1978mit..book.....B}
{Bateman}, G. 1978, {MHD instabilities}

\bibitem[{{Bian} {et~al.}(2022{\natexlab{a}}){Bian}, {Jiang}, {Feng}, {Zuo}, \&
  {Wang}}]{2022ApJ...925L...7B}
{Bian}, X., {Jiang}, C., {Feng}, X., {Zuo}, P., \& {Wang}, Y.
  2022{\natexlab{a}}, \apjl, 925, L7, \dodoi{10.3847/2041-8213/ac4980}

\bibitem[{{Bian} {et~al.}(2021){Bian}, {Jiang}, {Feng}, {Zuo}, {Wang}, \&
  {Wang}}]{2021arXiv211104984B}
{Bian}, X., {Jiang}, C., {Feng}, X., {et~al.} 2021, arXiv e-prints,
  arXiv:2111.04984.
\newblock \doarXiv{2111.04984}

\bibitem[{{Bian} {et~al.}(2022{\natexlab{b}}){Bian}, {Jiang}, {Feng}, {Zuo},
  {Wang}, \& {Wang}}]{2022A&A...658A.174B}
---. 2022{\natexlab{b}}, \aap, 658, A174, \dodoi{10.1051/0004-6361/202141996}

\bibitem[{{Brown} {et~al.}(2003){Brown}, {Nightingale}, {Alexander},
  {Schrijver}, {Metcalf}, {Shine}, {Title}, \& {Wolfson}}]{2003SoPh..216...79B}
{Brown}, D.~S., {Nightingale}, R.~W., {Alexander}, D., {et~al.} 2003, \solphys,
  216, 79, \dodoi{10.1023/A:1026138413791}

\bibitem[{{Carmichael}(1964)}]{1964NASSP..50..451C}
{Carmichael}, H. 1964, {A Process for Flares}, Vol.~50, 451

\bibitem[{{Chang} \& {Tot}(1993)}]{1993LNP...414..396C}
{Chang}, S.~C., \& {Tot}, W.~M. 1993, {A brief description of a new numerical
  framework for solving conservation laws {\textemdash} The method of
  space-time conservation element and solution element}, ed. M.~{Napolitano} \&
  F.~{Sabetta}, Vol. 414, 396--400, \dodoi{10.1007/3-540-56394-6\_255}

\bibitem[{{Chen}(2011)}]{2011LRSP....8....1C}
{Chen}, P.~F. 2011, Living Reviews in Solar Physics, 8, 1,
  \dodoi{10.12942/lrsp-2011-1}

\bibitem[{{DeLuca} {et~al.}(2005){DeLuca}, {Weber}, {Sette}, {Golub},
  {Shibasaki}, {Sakao}, \& {Kano}}]{2005AdSpR..36.1489D}
{DeLuca}, E.~E., {Weber}, M.~A., {Sette}, A.~L., {et~al.} 2005, Advances in
  Space Research, 36, 1489, \dodoi{10.1016/j.asr.2004.12.073}

\bibitem[{{Duan} {et~al.}(2019){Duan}, {Jiang}, {He}, {Feng}, {Zou}, \&
  {Cui}}]{2019ApJ...884...73D}
{Duan}, A., {Jiang}, C., {He}, W., {et~al.} 2019, \apj, 884, 73,
  \dodoi{10.3847/1538-4357/ab3e33}

\bibitem[{{Evershed}(1909)}]{1909MNRAS..69..454E}
{Evershed}, J. 1909, \mnras, 69, 454, \dodoi{10.1093/mnras/69.5.454}

\bibitem[{{Fan}(2005)}]{2005ApJ...630..543F}
{Fan}, Y. 2005, \apj, 630, 543, \dodoi{10.1086/431733}

\bibitem[{{Fan}(2011)}]{2011ApJ...740...68F}
---. 2011, \apj, 740, 68, \dodoi{10.1088/0004-637X/740/2/68}

\bibitem[{{Forbes}(2000)}]{2000JGR10523153F}
{Forbes}, T.~G. 2000, \jgr, 105, 23153, \dodoi{10.1029/2000JA000005}

\bibitem[{{Forbes} \& {Isenberg}(1991)}]{1991ApJ...373..294F}
{Forbes}, T.~G., \& {Isenberg}, P.~A. 1991, \apj, 373, 294,
  \dodoi{10.1086/170051}

\bibitem[{{Golub} {et~al.}(2007){Golub}, {Deluca}, {Austin}, {Bookbinder},
  {Caldwell}, {Cheimets}, {Cirtain}, {Cosmo}, {Reid}, {Sette}, {Weber},
  {Sakao}, {Kano}, {Shibasaki}, {Hara}, {Tsuneta}, {Kumagai}, {Tamura},
  {Shimojo}, {McCracken}, {Carpenter}, {Haight}, {Siler}, {Wright}, {Tucker},
  {Rutledge}, {Barbera}, {Peres}, \& {Varisco}}]{2007SoPh..243...63G}
{Golub}, L., {Deluca}, E., {Austin}, G., {et~al.} 2007, \solphys, 243, 63,
  \dodoi{10.1007/s11207-007-0182-1}

\bibitem[{{Guo} {et~al.}(2008){Guo}, {Ding}, {Wiegelmann}, \&
  {Li}}]{2008ApJ...679.1629G}
{Guo}, Y., {Ding}, M.~D., {Wiegelmann}, T., \& {Li}, H. 2008, \apj, 679, 1629,
  \dodoi{10.1086/587684}

\bibitem[{{Guo} {et~al.}(2021){Guo}, {Zhong}, {Ding}, {Chen}, {Xia}, \&
  {Keppens}}]{2021ApJ...919...39G}
{Guo}, Y., {Zhong}, Z., {Ding}, M.~D., {et~al.} 2021, \apj, 919, 39,
  \dodoi{10.3847/1538-4357/ac10c8}

\bibitem[{{Hirayama}(1974)}]{1974SoPh...34..323H}
{Hirayama}, T. 1974, \solphys, 34, 323, \dodoi{10.1007/BF00153671}

\bibitem[{{Inoue} {et~al.}(2014){Inoue}, {Hayashi}, {Magara}, {Choe}, \&
  {Park}}]{2014ApJ...788..182I}
{Inoue}, S., {Hayashi}, K., {Magara}, T., {Choe}, G.~S., \& {Park}, Y.~D. 2014,
  \apj, 788, 182, \dodoi{10.1088/0004-637X/788/2/182}

\bibitem[{{Inoue} {et~al.}(2010){Inoue}, {Kusano}, \&
  {Magara}}]{2010AGUFMSH31A1785I}
{Inoue}, S., {Kusano}, K., \& {Magara}, T. 2010, in AGU Fall Meeting Abstracts,
  Vol. 2010, SH31A--1785

\bibitem[{{Jiang} {et~al.}(2021{\natexlab{a}}){Jiang}, {Bian}, {Sun}, \&
  {Feng}}]{2021FrP.....9..224J}
{Jiang}, C., {Bian}, X., {Sun}, T., \& {Feng}, X. 2021{\natexlab{a}}, Frontiers
  in Physics, 9, 224, \dodoi{10.3389/fphy.2021.646750}

\bibitem[{Jiang {et~al.}(2022{\natexlab{a}})Jiang, Feng, Guo, \&
  Hu}]{JIANG2022100236}
Jiang, C., Feng, X., Guo, Y., \& Hu, Q. 2022{\natexlab{a}}, The Innovation, 3,
  100236, \dodoi{https://doi.org/10.1016/j.xinn.2022.100236}

\bibitem[{{Jiang} {et~al.}(2018){Jiang}, {Feng}, \& {Hu}}]{2018ApJ...866...96J}
{Jiang}, C., {Feng}, X., \& {Hu}, Q. 2018, \apj, 866, 96,
  \dodoi{10.3847/1538-4357/aadd08}

\bibitem[{{Jiang} {et~al.}(2010){Jiang}, {Feng}, {Zhang}, \&
  {Zhong}}]{2010SoPh..267..463J}
{Jiang}, C., {Feng}, X., {Zhang}, J., \& {Zhong}, D. 2010, \solphys, 267, 463,
  \dodoi{10.1007/s11207-010-9649-6}

\bibitem[{{Jiang} \& {Hu}(2018)}]{2018shin.confE.208J}
{Jiang}, C., \& {Hu}, Q. 2018, in Solar Heliospheric and INterplanetary
  Environment (SHINE 2018), 208

\bibitem[{{Jiang} {et~al.}(2016{\natexlab{a}}){Jiang}, {Wu}, {Feng}, \&
  {Hu}}]{2016NatCo...711522J}
{Jiang}, C., {Wu}, S.~T., {Feng}, X., \& {Hu}, Q. 2016{\natexlab{a}}, Nature
  Communications, 7, 11522, \dodoi{10.1038/ncomms11522}

\bibitem[{{Jiang} {et~al.}(2016{\natexlab{b}}){Jiang}, {Wu}, {Yurchyshyn},
  {Wang}, {Feng}, \& {Hu}}]{2016ApJ...828...62J}
{Jiang}, C., {Wu}, S.~T., {Yurchyshyn}, V., {et~al.} 2016{\natexlab{b}}, \apj,
  828, 62, \dodoi{10.3847/0004-637X/828/1/62}

\bibitem[{{Jiang} {et~al.}(2021{\natexlab{b}}){Jiang}, {Feng}, {Liu}, {Yan},
  {Hu}, {Moore}, {Duan}, {Cui}, {Zuo}, {Wang}, \& {Wei}}]{2021NatAs...5.1126J}
{Jiang}, C., {Feng}, X., {Liu}, R., {et~al.} 2021{\natexlab{b}}, Nature
  Astronomy, 5, 1126, \dodoi{10.1038/s41550-021-01414-z}

\bibitem[{Jiang {et~al.}(2022{\natexlab{b}})Jiang, feng, Bian, Zou, Duan, Yan,
  Hu, He, Zuo, \& Wang}]{pub.1147837699}
Jiang, C., feng, x., Bian, X., {et~al.} 2022{\natexlab{b}}, Research Square,
  \dodoi{10.21203/rs.3.rs-1613867/v1}

\bibitem[{{Jing} {et~al.}(2008){Jing}, {Wiegelmann}, {Suematsu}, {Kubo}, \&
  {Wang}}]{2008ApJ...676L..81J}
{Jing}, J., {Wiegelmann}, T., {Suematsu}, Y., {Kubo}, M., \& {Wang}, H. 2008,
  \apjl, 676, L81, \dodoi{10.1086/587058}

\bibitem[{{Jing} {et~al.}(2021){Jing}, {Inoue}, {Lee}, {Li}, {Nita}, {Xu},
  {Liu}, {Gary}, \& {Wang}}]{2021ApJ...922..108J}
{Jing}, J., {Inoue}, S., {Lee}, J., {et~al.} 2021, \apj, 922, 108,
  \dodoi{10.3847/1538-4357/ac26c7}

\bibitem[{{Karpen} {et~al.}(2012){Karpen}, {Antiochos}, \&
  {DeVore}}]{2012ApJ...760...81K}
{Karpen}, J.~T., {Antiochos}, S.~K., \& {DeVore}, C.~R. 2012, \apj, 760, 81,
  \dodoi{10.1088/0004-637X/760/1/81}

\bibitem[{{Kliem} \& {T{\"o}r{\"o}k}(2006)}]{2006PhRvL..96y5002K}
{Kliem}, B., \& {T{\"o}r{\"o}k}, T. 2006, \prl, 96, 255002,
  \dodoi{10.1103/PhysRevLett.96.255002}

\bibitem[{{Kopp} \& {Pneuman}(1976)}]{1976SoPh...50...85K}
{Kopp}, R.~A., \& {Pneuman}, G.~W. 1976, \solphys, 50, 85,
  \dodoi{10.1007/BF00206193}

\bibitem[{{Kosugi} {et~al.}(2007){Kosugi}, {Matsuzaki}, {Sakao}, {Shimizu},
  {Sone}, {Tachikawa}, {Hashimoto}, {Minesugi}, {Ohnishi}, {Yamada}, {Tsuneta},
  {Hara}, {Ichimoto}, {Suematsu}, {Shimojo}, {Watanabe}, {Shimada}, {Davis},
  {Hill}, {Owens}, {Title}, {Culhane}, {Harra}, {Doschek}, \&
  {Golub}}]{2007SoPh..243....3K}
{Kosugi}, T., {Matsuzaki}, K., {Sakao}, T., {et~al.} 2007, \solphys, 243, 3,
  \dodoi{10.1007/s11207-007-9014-6}

\bibitem[{{Kubo} {et~al.}(2007){Kubo}, {Yokoyama}, {Katsukawa}, {Lites},
  {Tsuneta}, {Suematsu}, {Ichimoto}, {Shimizu}, {Nagata}, {Tarbell}, {Shine},
  {Title}, \& {Elmore David}}]{2007PASJ...59S.779K}
{Kubo}, M., {Yokoyama}, T., {Katsukawa}, Y., {et~al.} 2007, \pasj, 59, S779,
  \dodoi{10.1093/pasj/59.sp3.S779}

\bibitem[{{Kusano} {et~al.}(2012){Kusano}, {Bamba}, {Yamamoto}, {Iida},
  {Toriumi}, \& {Asai}}]{2012ApJ...760...31K}
{Kusano}, K., {Bamba}, Y., {Yamamoto}, T.~T., {et~al.} 2012, \apj, 760, 31,
  \dodoi{10.1088/0004-637X/760/1/31}

\bibitem[{{Lin} \& {Forbes}(2000)}]{2000JGR...105.2375L}
{Lin}, J., \& {Forbes}, T.~G. 2000, \jgr, 105, 2375,
  \dodoi{10.1029/1999JA900477}

\bibitem[{{Liu} {et~al.}(2016){Liu}, {Kliem}, {Titov}, {Chen}, {Wang}, {Wang},
  {Liu}, {Xu}, \& {Wiegelmann}}]{2016ApJ...818..148L}
{Liu}, R., {Kliem}, B., {Titov}, V.~S., {et~al.} 2016, \apj, 818, 148,
  \dodoi{10.3847/0004-637X/818/2/148}

\bibitem[{{MacNeice} {et~al.}(2000){MacNeice}, {Olson}, {Mobarry}, {de
  Fainchtein}, \& {Packer}}]{2000CoPhC.126..330M}
{MacNeice}, P., {Olson}, K.~M., {Mobarry}, C., {de Fainchtein}, R., \&
  {Packer}, C. 2000, Computer Physics Communications, 126, 330,
  \dodoi{10.1016/S0010-4655(99)00501-9}

\bibitem[{{Magara} \& {Tsuneta}(2008)}]{2008PASJ...60.1181M}
{Magara}, T., \& {Tsuneta}, S. 2008, \pasj, 60, 1181,
  \dodoi{10.1093/pasj/60.5.1181}

\bibitem[{{Min} \& {Chae}(2009)}]{2009SoPh..258..203M}
{Min}, S., \& {Chae}, J. 2009, \solphys, 258, 203,
  \dodoi{10.1007/s11207-009-9425-7}

\bibitem[{{Moore} \& {Labonte}(1980)}]{1980IAUS...91..207M}
{Moore}, R.~L., \& {Labonte}, B.~J. 1980, in Solar and Interplanetary Dynamics,
  ed. M.~{Dryer} \& E.~{Tandberg-Hanssen}, Vol.~91, 207--210

\bibitem[{{Moore} \& {Roumeliotis}(1992)}]{1992LNP...399...69M}
{Moore}, R.~L., \& {Roumeliotis}, G. 1992, {Triggering of Eruptive Flares -
  Destabilization of the Preflare Magnetic Field Configuration}, ed.
  Z.~{Svestka}, B.~V. {Jackson}, \& M.~E. {Machado}, Vol. 399, 69,
  \dodoi{10.1007/3-540-55246-4\_79}

\bibitem[{{Moore} {et~al.}(2001){Moore}, {Sterling}, {Hudson}, \&
  {Lemen}}]{2001ApJ...552..833M}
{Moore}, R.~L., {Sterling}, A.~C., {Hudson}, H.~S., \& {Lemen}, J.~R. 2001,
  \apj, 552, 833, \dodoi{10.1086/320559}

\bibitem[{{Muhamad} {et~al.}(2017){Muhamad}, {Kusano}, {Inoue}, \&
  {Shiota}}]{2017ApJ...842...86M}
{Muhamad}, J., {Kusano}, K., {Inoue}, S., \& {Shiota}, D. 2017, \apj, 842, 86,
  \dodoi{10.3847/1538-4357/aa750e}

\bibitem[{{Petschek}(1964)}]{1964NASSP..50..425P}
{Petschek}, H.~E. 1964, {Magnetic Field Annihilation}, Vol.~50, 425

\bibitem[{{Prasad} {et~al.}(2017){Prasad}, {Bhattacharyya}, \&
  {Kumar}}]{2017ApJ...840...37P}
{Prasad}, A., {Bhattacharyya}, R., \& {Kumar}, S. 2017, \apj, 840, 37,
  \dodoi{10.3847/1538-4357/aa6c58}

\bibitem[{{Ravindra} \& {Howard}(2010)}]{2010BASI...38..147R}
{Ravindra}, B., \& {Howard}, T.~A. 2010, Bulletin of the Astronomical Society
  of India, 38, 147.
\newblock \doarXiv{1010.5849}

\bibitem[{{Ravindra} {et~al.}(2011){Ravindra}, {Venkatakrishnan}, {Tiwari}, \&
  {Bhattacharyya}}]{2011ApJ...740...19R}
{Ravindra}, B., {Venkatakrishnan}, P., {Tiwari}, S.~K., \& {Bhattacharyya}, R.
  2011, \apj, 740, 19, \dodoi{10.1088/0004-637X/740/1/19}

\bibitem[{{Schmieder} {et~al.}(2013){Schmieder}, {D{\'e}moulin}, \&
  {Aulanier}}]{2013AdSpR..51.1967S}
{Schmieder}, B., {D{\'e}moulin}, P., \& {Aulanier}, G. 2013, Advances in Space
  Research, 51, 1967, \dodoi{10.1016/j.asr.2012.12.026}

\bibitem[{{Schrijver} {et~al.}(2008){Schrijver}, {DeRosa}, {Metcalf}, {Barnes},
  {Lites}, {Tarbell}, {McTiernan}, {Valori}, {Wiegelmann}, {Wheatland},
  {Amari}, {Aulanier}, {D{\'e}moulin}, {Fuhrmann}, {Kusano}, {R{\'e}gnier}, \&
  {Thalmann}}]{2008ApJ...675.1637S}
{Schrijver}, C.~J., {DeRosa}, M.~L., {Metcalf}, T., {et~al.} 2008, \apj, 675,
  1637, \dodoi{10.1086/527413}

\bibitem[{{Sturrock}(1966)}]{1966Natur.211..695S}
{Sturrock}, P.~A. 1966, \nat, 211, 695, \dodoi{10.1038/211695a0}

\bibitem[{{Sturrock}(1991)}]{1991ApJ...380..655S}
---. 1991, \apj, 380, 655, \dodoi{10.1086/170620}

\bibitem[{{Su} {et~al.}(2007){Su}, {Golub}, {van Ballegooijen}, {Deluca},
  {Reeves}, {Sakao}, {Kano}, {Narukage}, \& {Shibasaki
  Kiyoto}}]{2007PASJ...59S.785S}
{Su}, Y., {Golub}, L., {van Ballegooijen}, A., {et~al.} 2007, \pasj, 59, S785,
  \dodoi{10.1093/pasj/59.sp3.S785}

\bibitem[{{Tan} {et~al.}(2009){Tan}, {Chen}, {Abramenko}, \&
  {Wang}}]{2009ApJ...690.1820T}
{Tan}, C., {Chen}, P.~F., {Abramenko}, V., \& {Wang}, H. 2009, \apj, 690, 1820,
  \dodoi{10.1088/0004-637X/690/2/1820}

\bibitem[{{Titov} {et~al.}(2002){Titov}, {Hornig}, \&
  {D{\'e}moulin}}]{2002JGRA..107.1164T}
{Titov}, V.~S., {Hornig}, G., \& {D{\'e}moulin}, P. 2002, Journal of
  Geophysical Research (Space Physics), 107, 1164, \dodoi{10.1029/2001JA000278}

\bibitem[{{T{\"o}r{\"o}k} \& {Kliem}(2005)}]{2005ApJ...630L..97T}
{T{\"o}r{\"o}k}, T., \& {Kliem}, B. 2005, \apjl, 630, L97,
  \dodoi{10.1086/462412}

\bibitem[{{T{\"o}r{\"o}k} {et~al.}(2004){T{\"o}r{\"o}k}, {Kliem}, \&
  {Titov}}]{2004A&A...413L..27T}
{T{\"o}r{\"o}k}, T., {Kliem}, B., \& {Titov}, V.~S. 2004, \aap, 413, L27,
  \dodoi{10.1051/0004-6361:20031691}

\bibitem[{{T{\"o}r{\"o}k} {et~al.}(2013){T{\"o}r{\"o}k}, {Temmer}, {Valori},
  {Veronig}, {van Driel-Gesztelyi}, \& {Vr{\v{s}}nak}}]{2013SoPh..286..453T}
{T{\"o}r{\"o}k}, T., {Temmer}, M., {Valori}, G., {et~al.} 2013, \solphys, 286,
  453, \dodoi{10.1007/s11207-013-0269-9}

\bibitem[{{Tsuneta} {et~al.}(2008){Tsuneta}, {Ichimoto}, {Katsukawa}, {Nagata},
  {Otsubo}, {Shimizu}, {Suematsu}, {Nakagiri}, {Noguchi}, {Tarbell}, {Title},
  {Shine}, {Rosenberg}, {Hoffmann}, {Jurcevich}, {Kushner}, {Levay}, {Lites},
  {Elmore}, {Matsushita}, {Kawaguchi}, {Saito}, {Mikami}, {Hill}, \&
  {Owens}}]{2008SoPh..249..167T}
{Tsuneta}, S., {Ichimoto}, K., {Katsukawa}, Y., {et~al.} 2008, \solphys, 249,
  167, \dodoi{10.1007/s11207-008-9174-z}

\bibitem[{{Vemareddy} {et~al.}(2012){Vemareddy}, {Ambastha}, \&
  {Maurya}}]{2012ApJ...761...60V}
{Vemareddy}, P., {Ambastha}, A., \& {Maurya}, R.~A. 2012, \apj, 761, 60,
  \dodoi{10.1088/0004-637X/761/1/60}

\bibitem[{{Wang} {et~al.}(2008){Wang}, {Jing}, {Tan}, {Wiegelmann}, \&
  {Kubo}}]{2008ApJ...687..658W}
{Wang}, H., {Jing}, J., {Tan}, C., {Wiegelmann}, T., \& {Kubo}, M. 2008, \apj,
  687, 658, \dodoi{10.1086/592082}

\bibitem[{{Yan} {et~al.}(2008){Yan}, {Qu}, \& {Kong}}]{2008MNRAS.391.1887Y}
{Yan}, X.~L., {Qu}, Z.~Q., \& {Kong}, D.~F. 2008, \mnras, 391, 1887,
  \dodoi{10.1111/j.1365-2966.2008.14002.x}

\bibitem[{{Yan} {et~al.}(2018){Yan}, {Wang}, {Pan}, {Kong}, {Xue}, {Yang},
  {Li}, \& {Feng}}]{2018ApJ...856...79Y}
{Yan}, X.~L., {Wang}, J.~C., {Pan}, G.~M., {et~al.} 2018, \apj, 856, 79,
  \dodoi{10.3847/1538-4357/aab153}

\bibitem[{{Yokoyama} \& {Shibata}(1994)}]{1994ApJ...436L.197Y}
{Yokoyama}, T., \& {Shibata}, K. 1994, \apjl, 436, L197, \dodoi{10.1086/187666}

\bibitem[{{Zhang} {et~al.}(2007){Zhang}, {Li}, \& {Song}}]{2007ApJ...662L..35Z}
{Zhang}, J., {Li}, L., \& {Song}, Q. 2007, \apjl, 662, L35,
  \dodoi{10.1086/519280}

\bibitem[{{Zhang} {et~al.}(2002){Zhang}, {Yu}, \&
  {Chang}}]{2002JCoPh.175..168Z}
{Zhang}, Z.-C., {Yu}, S.~T.~J., \& {Chang}, S.-C. 2002, Journal of
  Computational Physics, 175, 168, \dodoi{10.1006/jcph.2001.6934}

\end{thebibliography}
\bibliographystyle{aasjournal}

\end{document}